\documentclass[aps,prx,longbibliography,twocolumn,superscriptaddress]{revtex4-2}
\usepackage{amsfonts}
\usepackage{amsmath}
\usepackage{graphicx}
\usepackage{dsfont}
\usepackage{diagbox}
\usepackage{array}
\usepackage{amssymb}
\usepackage{bbm}
\usepackage{subfigure}
\usepackage{MnSymbol}
\usepackage{url}
\usepackage{times}
\usepackage{manfnt}
\usepackage{booktabs}
\usepackage{xcolor}
\usepackage{svg}
\definecolor{darkblue}{HTML}{004D6B}
\definecolor{darkred}{HTML}{8c1515}
\definecolor{darkgreen}{HTML}{006400}
\usepackage{hyperref}
\hypersetup{
    pdftitle={BornDQPT},
	colorlinks=true,
	urlcolor=darkred,
	citecolor=darkblue,
	linkcolor=darkred,
	breaklinks
}

\pagecolor{white}
\usepackage{ulem}
\usepackage{physics}
\usepackage{enumitem}
\usepackage{wasysym}
\usepackage{multirow}

\newcommand\redsout{\bgroup\markoverwith{\textcolor{red}{\rule[0.5ex]{2pt}{0.4pt}}}\ULon}

\begin{document}

\title{
	Born-rule statistical dynamical quantum phase transitions under measurement 
}

\author{Guan-Hua Chen}
\affiliation{The Hong Kong University of Science and Technology (Guangzhou), Nansha, Guangzhou, 511400, Guangdong, China}
\affiliation{Department of Physics, South China University of Technology, Guangzhou 510640, China}

\author{Guo-Yi Zhu}
\email{guoyizhu@hkust-gz.edu.cn}
\affiliation{The Hong Kong University of Science and Technology (Guangzhou), Nansha, Guangzhou, 511400, Guangdong, China}

\date{\today}
\begin{abstract}
Dynamical quantum phase transitions (DQPTs) occur at times when a quantum state exhibits a nonanalytic change in its return probability. 
This can be viewed as the probability of collapsing the evolved state to the initial state by quantum measurement. 
However, the initial wave function usually has exponentially small amplitude in the late time evolved state. 
Here we perform statistical characterization for all the possible post-measurement states distributed according to the Born's rule, by sampling a one-dimensional quantum Ising chain after a quantum quench dynamics. The statistical ensemble can also be viewed as a mixed state when the time evolved state is subjected to maximally dephasing noise in a certain basis. 
We map the distribution to a statistical model and characterize its effective ``energy" spectrum, and introduce the average dynamical free energy, establishing a framework for the {\it statistical} DQPTs. We show the recovering of DQPT under high-moment average and a delocalized level distribution following critical times. Through analytic continuation into the complex time plane, we demonstrate the vanishing of Yang-Lee-Fisher zeros and the emergent level crossing near critical times. Finally, we propose a measurement-based quantum computation protocol to simulate the unitary evolution via single-qubit measurements on a two-dimensional  cluster state. Our results provide a way for experimentally investigating statistical DQPTs in quantum devices, shedding light on the structured circuit sampling with insights from DQPT and generalizing the understanding of mixed state due to decoherence beyond equilibrium. 

\end{abstract}

\maketitle


{\it \bf Introduction.}--
Understanding nonequilibrium quantum dynamics in open systems stands as one of the central tasks in quantum simulation and information processing. 
In this regime, dynamical quantum phase transitions (DQPTs) extend the concept of equilibrium phase transitions into the time domain following a quantum quench~\cite{PhysRevLett.110.135704,Heyl_2018}. Within this analogy, the Loschmidt amplitude $\langle \psi_0|e^{-iHt}|\psi_0\rangle$ serves as a dynamical partition function whose Yang-Lee-Fisher (YLF) zeros~\cite{YangLee52i, YangLee52ii, fisher1965statistical} in the analytically  continued complex plane condense in the thermodynamic limit, resulting in nonanalytic kinks in the dynamical free energy~\cite{PhysRevB.87.195104,PhysRevB.89.125120,
PhysRevB.89.161105,PhysRevLett.115.140602,PhysRevB.93.104302,PhysRevB.93.144306,PhysRevB.96.134427,PhysRevLett.120.130601,Hamazaki2021,PhysRevB.104.075130,PhysRevLett.126.040602,PhysRevResearch.4.013250,rnv5-f32k,PhysRevLett.123.160603}. 
The conventional framework of DQPTs relies on the Loschmidt echo which quantifies the return probability to the initial state specifically, corresponding to a single trajectory in the exponentially large configuration space spanned by the time-evolved state~\cite{PhysRevLett.119.080501,Flaschner2018,PhysRevApplied.11.044080,PhysRevA.102.042222,sciadv.aba4935}. While this 
quantity captures nonanalyticities in the thermodynamic limit, it discards most of the remaining information of coherent superposition. 

Quantum measurements provide a universal and experimentally natural means to probe the neglected information in the evolved quantum state. According to the Born's rule, measuring the evolved state in a product basis yields a probabilistic distribution of bitstrings, where the Loschmidt echo corresponds to the probability of post-selecting the initial configuration. 
The measurement outcomes from repeated experiments form a statistical ensemble obeying the Born-rule probability distribution function. Such an ensemble for an unstructured quantum state following a random quantum circuit has been the central subject of the random circuit sampling problem~\cite{Lund2017,Bouland2019, Arute2019}, while here we focus on sampling the {\it structured} quantum state. This raises a critical question: beyond the single trajectory captured by the Loschmidt echo, does the statistical ensemble also exhibit DQPTs and if so, how to characterize such {\it statistical} DQPTs? Besides, the ensemble is also describable as a mixed state by subjecting the pure quantum state after a quantum quench to maximally dephasing noise, erasing all the off-diagonal elements of the density matrix. The nontrivial structure of the mixed state by decohering~\cite{LeeJianXu,FanBaoAltmanVishwanath,SalaGopalakrishnanOshikawaYou,EllisonCheng,WangWuWang,YHChenGrover,YHChenGroverSeparability,LessaStrongToWeak,WangKielyEtAl2026DephasedFermiGas,SohalPrem,SuYangJian,LeeExactCalculation,EcksteinPRX,DecoherenceWangEtAl,eckstein2025learningtransitionstopologicalsurface,DiehlEtAl2025,wan2025revisitingnishimorimulticriticalitylens} topological states or critical states~\cite{Garratt22,Garratt23measureising,Alicea23measureising,Alicea24teleportation,jian23measureising,Ludwig24measurecritical,liu2024boundary,LuitzGarratt2024MeasStatesHigherDim,hoshino2024entanglement,tang2024critical,PuetzGarrattNishimoriTrebstZhu2025,PatilPuetzTrebstZhuLudwig2026} has been an emerging active research direction recently, while here we go beyond the equilibrium ground state and decohere a time-evolved highly {\it dynamical} state out of equilibrium. Such a mixed state loses quantum coherence but maintains the population information of the dynamical quantum state, such as the diagonal entropy~\cite{POLKOVNIKOV2011486,ChenGrover2025Zipping}. Note the measurement-induced mixed state we consider here is distinct from the mixed-state DQPT~\cite{PhysRevB.96.180304,PhysRevB.96.180303,PhysRevB.98.134310} which considered the return probability of a (finite temperature) mixed state evolved by a unitary. The ``statistical'' DQPT in our model, if existing, captures the transition of this mixed state, as well as the diagonal entropy of the evolved quantum state.

\begin{figure*}[t]
	\centering
	\includegraphics[width=\textwidth]{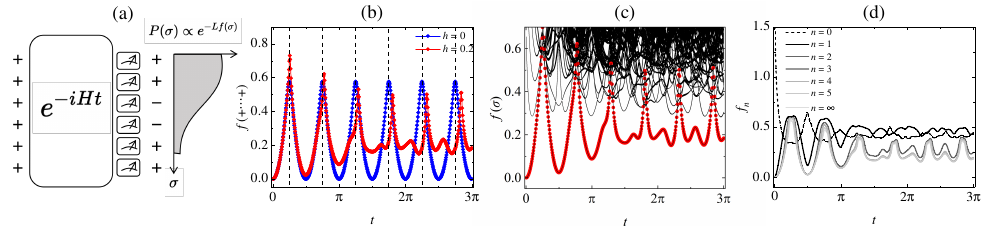}
	\caption{
	{\bf Dynamical free energy spectrum.} (a) Schematic of the model: a quantum quench dynamics by Ising model in ferromagnetic phase followed by measurements in $X$ basis. Repeating the protocol yields a Born-distributed ensemble of bitstrings $\sigma$, whose Born probability $P(\sigma)\propto e^{-Lf(\sigma)}$ defines the effective free energy $f(\sigma)$, conditioned upon $\sigma$ (the temporal boundary condition of spacetime path integral). (b) The evolution of post-selected dynamical free energy $f(+\cdots +)$ under PBC. The nonanalytic period $t^*$ without transverse field is exactly $\pi/2$ labeled by vertical dashed lines. 
	(c) The full energy spectrum $f(\sigma)$ for $h=0.2$. Red dots highlight $f(+\cdots+)$, which acts as the ``ground state" except near the critical times. 
	(d) Dynamical free energy $f_n$ averaged over Born probability at varying moments $n=0\sim5$ and $\infty$. When all measurement outcomes are collected according to Born's rule, the mixed state~\eqref{eq:mixed} can be viewed as occupying these ``energy'' levels at unit-temperature $n=1$. As one raises the moment $n$, the mixed state effectively goes to lower temperature, with $n=\infty$ corresponding to zero temperature. 
	Computation is performed by exactly computing Born's probability function for $L=12$ qubits. 
	}
	\label{fig:fnband} 
\end{figure*}

In this work, we investigate statistical DQPTs in the finite-size transverse-field Ising model by generalizing the Loschmidt echo to a weighted spectrum over all configurations. To describe the statistical ensemble, we introduce the average dynamical free energy, defined by the Born probability distribution from the measurement-induced mixed state. Utilizing tensor network~\cite{ITensor} and finite-size scaling, we reveal that statistical averages wash out nonanalytic characteristics of DQPTs, which can be recovered in larger systems. The statistical ensemble further exhibits a multifractality-to-extended crossover around the critical time, quantified by the scaling of participation entropy. In the complex time plane, we analyze the vanishing of YLF zeros and the level-crossing dynamics of the energy spectrum. Finally, we propose a measurement-based quantum computation (MBQC) protocol in a cluster state to simulate dynamics and identify the bulk-boundary correspondence in measurement-induced random circuits.

{\it \bf Model.}--
Our model setup consists of two parts: unitary evolution followed by measurements or dephasing noise. 
The DQPT~\cite{PhysRevLett.110.135704} was firstly introduced in the one-dimensional (1D) transverse-field Ising model with Pauli-$Z$ and Pauli-$X$ as
\begin{equation} \label{eqising}
	H=-J\sum_j Z_jZ_{j+1}-h\sum_j X_j\ ,
\end{equation}
whose ferromagnetic and paramagnetic phases are separated by the critical point $h = 1$. 
For the time evolution part, we consider the quantum quench from paramagnetic state $\ket{+\cdots +}=\ket{+}^{\otimes L}$ to the ferromagnetic phase, yielding the evolved state $ \ket{\psi} = e^{-iH t}\ket{+\cdots +}$. Concretely, in the following numerical simulation, we set $J=1$, $h=0.2$ under periodic boundary conditions (PBC) with time step $\Delta t = \pi/160$, total time $T=3\pi$, and system size $L=12$ unless otherwise specified.

Measuring a quantum system acts as noise that transforms a pure state into mixed and we consider the maximally dephased mixed state $\rho = \mathcal{N}(\ketbra{\psi}) $ with maximally dephasing noise $\mathcal{N}(\cdot) = [(\cdot) + X(\cdot)X]/2$, which can be realized by projective $X$ measurement. The mixed state is a diagonal density matrix in the $X$ basis with diagonal elements  $P(\sigma)$
\begin{equation} \label{eq:mixed}
    \rho = \sum_{\sigma} P(\sigma) \ketbra{\sigma} \ ,
\end{equation}
where $P(\sigma)=|\braket{\sigma}{\psi}|^2$ is the Born's probability defined by the square of overlap between the time-evolved state and a post-selected reference state $\ket{\sigma}$.
Note that in the fermionic basis (by Jordan-Wigner transformation), the time-evolved state is a (non-interacting) Gaussian fermionic state, but the dephasing noise renders the resulting mixed state $\rho$ non-Gaussian~\cite{WangKielyEtAl2026DephasedFermiGas, GuoZhu26fermion}~\footnote{
		A comment on the numerical method: the probability can be viewed as the amplitude of a time evolved Gaussian fermion state in the Fock space, by Jordan Wigner transforming the Ising Hamiltonian to the free fermion Hamiltonian in the parity even sector $\prod_j X_j=+1$, which maps the periodic boundary condition of the spin chain to the anti-periodic boundary condition of the fermion chain. Using fermion basis saves the need to numerically diagonalizing the Hamiltonian or trotterizing the Hamiltonian. However, the numerical computation is limited by enumerating the measurement outcome configurations, especially if we want to analyze the multifractal structure with small moment $q$ that emphasizes the less probable configurations. Since in fermion approach one stores an $2L$-by-$2L$ matrix that captures the Gaussian fermion state, one has to enumerate $2^L$ times of computation to calculate the whole probability distribution function. Similarly for tensor network approximation approach, where the spatial complexity of numerical computation is polynomial but we still require exponential time complexity to extract the whole distribution function. 
		Therefore for most calculation we adopt exact state evolution by a trotterized Hamiltonian, with exponential overhead in the spatial complexity but not in time complexity. We also leave a discussion of numerical sampling in the appendix. }. 
Within the experimental framework, each snapshot enables the sampling of a bitstring through measurement in the $X$ basis, and repeated sampling yields the corresponding $P(\sigma)$ of bitstrings $\sigma$, schematically shown in Fig.~\ref{fig:fnband}(a).  
Note that our spectrum resembles the Loschmidt echo amplitude spectrum introduced in Ref.~\cite{PhysRevB.105.174307}, but our overlap states $\bra{\sigma}$ are not required to be translationally invariant, which is natural in a quantum measurement or dephasing noise setting.
%
Due to the Ising symmetry $[H,\prod_jX_j] = 0$,  $\sigma$ with an odd number of $\ket{-}$ states (i.e., odd parity under $\prod_jX_j$) has strictly vanishing probability $P(\sigma) =0$. 

The dynamical free energy of the time-evolved quantum spin chain, conditioned upon measurement outcome $\sigma$, follows as 
\begin{equation}
	f(\sigma)=-\frac{1}{L}\ln P(\sigma)= -\frac{1}{L}\ln |\bra{\sigma}e^{-iH t} \ket{+\cdots+}|^2 \ .
	\label{eq:fsigma}
\end{equation}
Here $\sigma$ is viewed as an external field that sets the temporal boundary condition of the quantum spin chain in its spacetime path integral, which determines the free energy of the quantum chain. By tracing out the quantum degrees of freedom, $\sigma$ picks up their internal energy as $f(\sigma)$.
Then the measured bitstring $\sigma$ with lower $f(\sigma)$ occurs with higher probability.

{\it\bf Dynamical free energy spectrum.}--
In this section, we focus on the $\sigma$-dependent energy spectrum composed of all trajectories and their statistical average under different moments.
The free energy dynamics of all allowed bitstrings collectively constitute the energy spectrum in Fig.~\ref{fig:fnband}(c). 
The ground state energy mostly traces the conventional Loschmidt echo rate function~\cite{PhysRevLett.110.135704} with a relatively large gap separating the ``ground state'' $\ket{+\cdots+}$ from the ``excited states'', except at DQPT critical times where the ``gap'' closes level crossings occur, see Appendix~\ref{Energy level distribution} for a illustrative ground-state trajectory. At these points, the approximate degeneracy of numerous levels precludes extracting further information from the spectrum. 

In this statistical ensemble, we sum the dynamical free energy of all possible bitstrings with specific weights assigned by a moment factor $n$. The $n$-th moment average dynamical free energy $f_n$ is defined as
\begin{equation} \label{eq:fn}
	f_n=-\frac{1}{L}\frac{\sum_{\sigma} P^n (\sigma)\ln P(\sigma)}{\sum_{\sigma} P^n(\sigma)}\ = \frac{\sum_{\sigma} P^n (\sigma)f(\sigma)}{\sum_{\sigma} P^n(\sigma)},
\end{equation}
normalized by $\sum_{\sigma} P^n(\sigma)$. The weight factor $P^n(\sigma)$ as a generalized energy filter for the distribution of $\sigma$, highlights bitstrings with higher probability. As $n$ increases, the bitstrings with lower free energy gradually dominate.

However, our mixed state actually lies at ``unit temperature'', $f_1 = -\frac{1}{L} \sum_\sigma P(\sigma)\ln P(\sigma)$ as the Shannon entropy of the measurement records~\cite{ZabaloGullansWilsonVasseurLudwigGopalakrishnanHusePixley, DecoherenceWangEtAl, putz2025flownishimoriuniversalityweakly, PatilPuetzTrebstZhuLudwig2026}, since $P(\sigma)\propto e^{-f(\sigma)L}$. The $n$-th moment dynamical free energy density Eq.~\eqref{eq:fn} can be intuitively understood as the average energy of a state lying at temperature $1/n$. $n=0$ corresponds to the infinite-temperature limit where all configurations are equally probable, while $n=\infty$ corresponds to the zero-temperature ground state where only the most probable configuration is occupied. The nonzero-moment average induces statistical decoherence that the destructive interference among different $\sigma$ erases the nonanalytic signatures of DQPTs, see Fig.~\ref{fig:fnband}(d). 
For low moment $n=1$, the thermal broadening is strong enough to wash out the features of $f(+\cdots+)$. As the moment increases toward infinity, $f_n$ converges to post-selected $f(+\cdots+)$ except the critical regions where level crossings occur. In this finite-size system, the nonanalytic DQPT signatures are statistically washed out, while it can be recovered by increasing the size, see End Matter for numerical evidence of finite-size scaling.

Specifically, infinite-temperature $f_0$ exhibits a behavior opposite to that of other $f_n$, because some bitstrings with extremely high free energies dominate the dynamics. By contrast, their lower probabilities $P(\sigma)$ instead render them negligible for high-moment average. Following the Born’s probability at Eq.~\eqref{eq:mixed}, in a dynamical experiment with multiple copies, the mixed-state $P(\sigma)$ and $f_n$ can be obtained by sampling each qubit of the evolved state in the Pauli-$X$ basis. We numerically simulate this measurement sampling and show the difference between $f_0$ and $f_{1}$, see Appendix \ref{Born sampling}.

\begin{figure}[t]
	\centering
	\includegraphics[width=0.9\columnwidth]{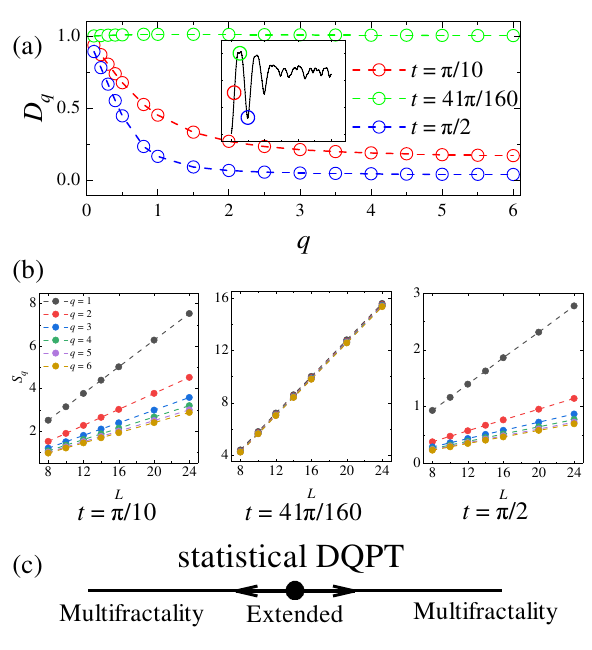}
	\caption{
		{\bf Delocalization at statistical DQPT}. 
		The statistical ensemble exhibits multifractality at generic times, while it becomes delocalized extended distribution at critical time of DQPT. 
		(a) Multifractal dimensions $D_q$ of the Born ensemble of snapshots as a function of moment index $q$ at three time slices before, at, and after a DQPT ($t= \pi/10, 41\pi/160, \pi/2$ labeled by colored circles, see the Inset), which are fitted by calculating PE $S_q$ in different system sizes at (b). Note that $q$ can take continuous non-integer values and the special case $q=1$ recovers the Shannon entropy: $S_1=-\sum_{\sigma}P(\sigma)\ln (\sigma)=f_1 L$. (d) Schematic crossover around the critical time.
		To reveal the complete structure of the Born's distribution, we employ exact enumeration of all measurement outcomes instead of Monte Carlo sampling here, for sizes up to $L=24$.
	}
	\label{tauvsq}
\end{figure}

To understand the distribution of energy levels across the DQPT, we examine the scaling of the participation entropy (PE) $S_q=\frac{1}{1-q}\ln \mathcal{Z}_q$. Here $\mathcal{Z}_q$ is the inverse participation ratio: $\mathcal{Z}_q = \sum_{\sigma} P^q(\sigma)$~\cite{Mirlin08rmp,PhysRevLett.123.180601,PhysRevA.111.052614} which can be viewed as the replicated partition function~\cite{Ludwig2020, DecoherenceWangEtAl} for $\sigma$. We find that for our time-evolved states $S_q$ always scales linearly with the system size as:  
\begin{equation} \label{eqipr}
	S_q \propto D_q L \ln 2  \ ,
\end{equation}
where $D_q$ denotes the multifractal dimension~\cite{Mirlin08rmp}. The fitted $D_q$ at given time slices are illustrated in Fig.~\ref{tauvsq}(a), with the finite-size linear scaling of PE in different indices $q$ shown in Fig.~\ref{tauvsq}(b), where their slopes are set by $D_q\ln2$ as Eq.~\ref{eqipr}. 
At generic times, the distribution $P(\sigma)$ exhibits multifractality, with continuously varying multifractal dimensions from $D_{q=0}=1$ to $D_{q\to \infty}\to 0$. Recalling that $q$ takes similar physical meaning as the inverse temperature, it can be naturally understood as the entropy density $S_q/L\propto D_q\ln 2$ continuously varies from $\ln 2$ at infinite temperature to $0$ at zero temperature by tuning $q$. However, at the DQPT location, $D_q\approx 1,\ \forall q$, which indicates a approximately delocalized extended distribution $P(\sigma)\approx 2^{-L}$ already at infinite temperature (up to a global constraint excluding the parity odd configurations, and the non-degeneracy structure of certain configuration in Fig.~\ref{fig:levelinversion}).
The above analysis provides a statistical perspective on DQPTs: at the critical time, the gap closing is accompanied by a condensation of energy levels into an extended phase with extensive degeneracy. After crossing the statistical DQPT, the system reenters the multifractal regime, completing a multifractality–extended–multifractality crossover, as shown in Fig.~\ref{tauvsq}(c).
We further provide the spectrum of single-shot $f(\sigma)$ at given time slices to show the level distribution, see Appendix \ref{Energy level distribution}.

\begin{figure}[t]
	\centering
	\includegraphics[width=\columnwidth]{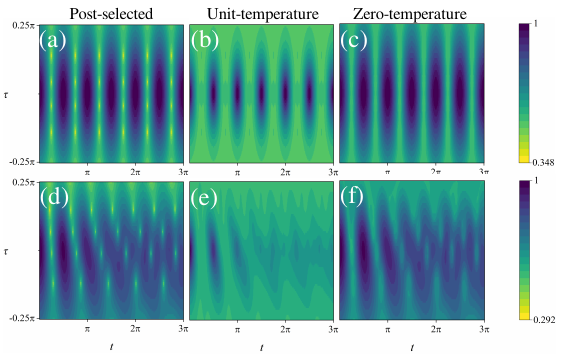}
	\caption{
	{\bf Statistically erased zeros in the complex time plane} with field $h=0$ (upper panel) and $h=0.2$ (lower panel).
	Finite-size ($L=12$) evolution of $e^{-f(+\cdots +)}$ and $e^{-f_n}$ in the complex time plane $\tau\in[-\pi/4, \pi/4]$. (a), (d) Post-selected $e^{-f(+\cdots +)} = P^{1/L}(+\cdots+)$ : Periodic zeros discretely lie on the complex plane (not the real axis $\tau=0$ as the critical real time in Fig.~\ref{fig:fnband}(b)). (b), (e) Unit-temperature $e^{-f_1}$: Discrete zeros at $\tau\ne 0$ and nonanalytic kinks on the real axis are both washed out by averaging. (c), (f) Ground-state $e^{-f_\infty}$: Discrete zeros in the complex plane are covered by lower-energy configurations. Although the finite-size level crossing still remains in the plane, sufficiently large size will recover the post-selected nonanalyticity, see End matter. Note that we cannot reach exact $e^{-f(+\cdots +)}=0$ due to the finite resolution.
	}
	\label{fig:complexpbc}
	\end{figure}

{\it \bf Fate of YLF zeros in statistical average.}--
A hallmark of the DQPT, generalizing the conventional phase transitions, is the presence~\cite{Heyl_2018,PhysRevLett.123.160603} of YLF zeros in the analytically continued complex plane of the tuning parameter. Here we analytically continue the real time $t$ into complex $t\to t+i\tau$, and investigate the YLF zeros of our statistical partition function. As a warmup let us consider the post-selected trajectory $P(+\cdots +)$ which recovers the conventional DQPT~\cite{Heyl_2018,PhysRevLett.101.120603,PhysRevB.87.195104}.
The post-selected partition function density $P(+\cdots +)^{1/L}$ is shown in Fig.~\ref{fig:complexpbc}(a) and (d). The hot spots signal the YLF zeros, which for the special case of $h=0$ agrees with the analytically solved~\cite{PhysRevLett.101.120603,PhysRevB.87.195104} locations (Eq.~\ref{eqzeros} in End Matter). The zeros are off the real axis of the complex time plane with a finite-size gap, but they condense in thermodynamic limit $L\to\infty$ closing the gap and triggering the transition along the real time axis~\cite{YangLee52i, YangLee52ii, fisher1965statistical}. 

Now we address the potential YLF zeros in the Born-rule statistical ensemble $\{P(\sigma)\}$. Note that the zeros of partition function correspond to divergence of free energy, for example $P(+\cdots +)=0\leftrightarrow f(+\cdots +)=\infty$. For the statistical ensemble let us consider the average free energy Eq.~\eqref{eq:fn} with $n=1$ i.e., the Shannon entropy density of the measurement record. $e^{-f_1}$ is shown in Fig.~\ref{fig:complexpbc}(b) and (e), which is lower-bounded by $e^{-\ln 2}$ and cannot vanish to exact zero. Therefore, YLF zeros, existing for certain post-selected states, disappear upon statistical average. This is because for those configurations that yield divergent free energy $-\ln P(\sigma)=\infty$, their Born probability also simultaneously vanishes $P(\sigma)=0$, and thus do not contribute to the Shannon entropy $-P(\sigma)\ln P(\sigma)$. From a different perspective, one can intuitively understand this as the mean energy of a mixed state at unit temperature cannot diverge. As the zero-temperature limit resembles the post-selected state, we also show $e^{-f_{\infty}}$ as in Fig.~\ref{fig:complexpbc}(c) and (f), which recovers the general feature of the partition function density as in post-selected cases (a) and (d). However, the sharp zeros again appear to disappear for our finite-size calculation. We find that this is because of a finite-size level inversion effect: $\ket{+\cdots +}$ at DQPT critical times has higher energy than the other configurations, but it becomes the ground state when extrapolated to the thermodynamic limit, see Fig.~\ref{fig:levelinversion}. Thus we expect that $e^{-f_{\infty}}$ in thermodynamic limit recovers the post-selected scenario with condensed YLF zeros. See Appendix \ref{Time slices in complex plane} for detailed $e^{-f_n}$ at typical time slices.  
The erasing of YLF zeros under Born averaging reflects a fundamental change in the analytic structure of the statistical partition function, due to the effective finite-temperature mixing. Thus one can increase $n$ for the statistical ensemble to lower the effective temperature so as to probe the zero-temperature DQPT as shown in Fig.~\ref{fig:fnband}(d). Numerical results under open boundary condition are provided as an expansion, see Appendix~\ref{Dynamical free energy under OBC}

\begin{figure}[hb]
	\centering
	\includegraphics[width=0.9\linewidth]{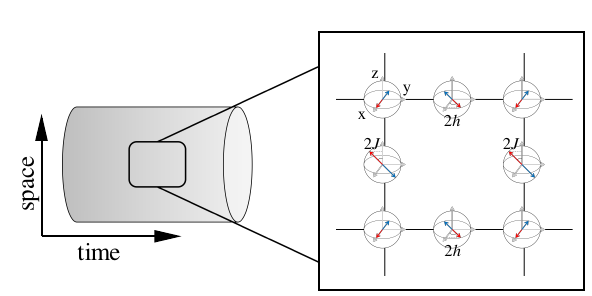}
	\caption{
	{\bf Bulk measurement protocol.}
	Schematic representation of a 2D cluster state on the Lieb lattice, whose spatial periodic boundary is defined on a cylinder (left). The corresponding time evolution operator of one Trotter step is $U(\Delta t)=e^{ihX \Delta t}e^{iJZZ \Delta t}$ whose unit cell of projective measurement on a plaquette is shown on the right. Measurement bases are represented on the Bloch spheres as $\cos{(\theta/2)}\ket{0}+ e^{i\phi}\sin(\theta/2)\ket{1}$ and positive (negative) measurement outcomes are denoted by red (blue) arrows. Corresponding angles related to the Ising model ignoring the time step are labeled beside arrows (mentioned as $(\theta,\phi)$ in the text).
	}
	\label{fmeasure} 
\end{figure}

{\it \bf Measurement-only protocol}.--
The whole protocol, including the unitary evolution part, can be realized by measuring a 2D state with boundary~\cite{Briegel2009,annurev020911125041, DecoherenceWangEtAl,PhysRevB.106.144311,PhysRevResearch.5.043069,PhysRevB.111.024312,SciPostPhys.14.5.129,PhysRevResearch.6.043018,PhysRevB.109.054432}. Then the correlated distribution $P(\sigma)$ we characterize corresponds to the marginal probability distribution of the boundary of a measured or dephased 2D state. 
Concretely, the unitary evolution from transverse-field Ising Hamiltonian can be mapped~\cite{stollenwerk2024,PhysRevB.111.024312} to 
measuring the bulk of a 2D cluster state, where the measurement angles follow the pattern shown in Fig.~\ref{fmeasure}. 
Here, the measurements in the 2D bulk introduce additional Born-rule randomness~\cite{DecoherenceWangEtAl}, which can in principle be corrected~\cite{Briegel2009}. 
Without active feedback correction, each bulk measurement outcome corresponds to distinct unitary operations. We leave a discussion of such situation in Appendix \ref{Measurement-based random circuits}.  

{\it \bf Outlook.}--
In summary, we have demonstrated the Born-rule statistical structure of DQPTs, quantified by the statistical average dynamical free energy. 
The moment factor $n$ as an effective inverse temperature allows a continuous crossover from a high-temperature mixed state to the zero-temperature ground state, where the post-selected nonanalytic behavior can be recovered. Our dynamics reveals the time-dependent redistribution of energy levels, where the multifractal distribution becomes delocalized into extended distribution at critical times of DQPT. The statistical transition of the Born ensemble complements the single-trajectory nonanalyticity of the conventional DQPT. 
On the other hand, in our analytic continuation $t\to t+i\tau$, YLF zeros are washed out by the Born-rule statistical average. This is because in view of $\sigma$ its Boltzmann weight $P(\sigma)\propto \bra{\psi(t+i\tau)}\ket{\sigma}\bra{\sigma}\ket{\psi(t+i\tau)}\geq 0$ is guaranteed to be positive semi-definite, even though the quantum chain evolution operator has been under analytical continued into complex plane. Therefore the YLF zeros could be located by further analytic continuation, for example by considering $P'(\sigma)\propto \bra{\psi(t-i\tau)}\ket{\sigma}\bra{\sigma}\ket{\psi(t+i\tau)}$ which is a complex probability. We leave this to future study. 

Our results provide several promising directions for future research. First, extending our sampling-based framework to non-integrable models or higher-dimensional systems remains an open challenge, where the energy spectrum may exhibit richer multifractal properties. Secondly, implementing the 
sampling scheme on current quantum hardware would be a natural next step, allowing its robustness against real experimental noise and decoherence to be tested. Finally, the measurement-based protocol offers a fresh perspective on relating certain higher-dimensional quantum state decoherence to the statistical DQPT.  
More broadly, this bulk-boundary measurement scenario is not limited to DQPTs but holds applicability to the investigation of general many-body dynamics and the preparation of specific entangled states.
\\

{\it \bf Data availability.}--
The numerical data shown in the figures are available on
Zenodo~\cite{Zenodo_data}.

\begin{acknowledgments}
	We acknowledge the support of National Natural Science Foundation of China - Young Scientists Fund (grant no.~12504181), Start-up Fund of HKUST(GZ) (grant no.~G0101000221), Guangdong provincial project (grant no.~2024QN11X201) and Guangdong Basic and Applied Basic Research Foundation (grant no.~2026A1515010965). 
\end{acknowledgments}

\bibliography{ref}

\begin{thebibliography}{82}%
\makeatletter
\providecommand \@ifxundefined [1]{%
 \@ifx{#1\undefined}
}%
\providecommand \@ifnum [1]{%
 \ifnum #1\expandafter \@firstoftwo
 \else \expandafter \@secondoftwo
 \fi
}%
\providecommand \@ifx [1]{%
 \ifx #1\expandafter \@firstoftwo
 \else \expandafter \@secondoftwo
 \fi
}%
\providecommand \natexlab [1]{#1}%
\providecommand \enquote  [1]{``#1''}%
\providecommand \bibnamefont  [1]{#1}%
\providecommand \bibfnamefont [1]{#1}%
\providecommand \citenamefont [1]{#1}%
\providecommand \href@noop [0]{\@secondoftwo}%
\providecommand \href [0]{\begingroup \@sanitize@url \@href}%
\providecommand \@href[1]{\@@startlink{#1}\@@href}%
\providecommand \@@href[1]{\endgroup#1\@@endlink}%
\providecommand \@sanitize@url [0]{\catcode `\\12\catcode `\$12\catcode
  `\&12\catcode `\#12\catcode `\^12\catcode `\_12\catcode `\%12\relax}%
\providecommand \@@startlink[1]{}%
\providecommand \@@endlink[0]{}%
\providecommand \url  [0]{\begingroup\@sanitize@url \@url }%
\providecommand \@url [1]{\endgroup\@href {#1}{\urlprefix }}%
\providecommand \urlprefix  [0]{URL }%
\providecommand \Eprint [0]{\href }%
\providecommand \doibase [0]{https://doi.org/}%
\providecommand \selectlanguage [0]{\@gobble}%
\providecommand \bibinfo  [0]{\@secondoftwo}%
\providecommand \bibfield  [0]{\@secondoftwo}%
\providecommand \translation [1]{[#1]}%
\providecommand \BibitemOpen [0]{}%
\providecommand \bibitemStop [0]{}%
\providecommand \bibitemNoStop [0]{.\EOS\space}%
\providecommand \EOS [0]{\spacefactor3000\relax}%
\providecommand \BibitemShut  [1]{\csname bibitem#1\endcsname}%
\let\auto@bib@innerbib\@empty
\bibitem [{\citenamefont {Heyl}\ \emph {et~al.}(2013)\citenamefont {Heyl},
  \citenamefont {Polkovnikov},\ and\ \citenamefont
  {Kehrein}}]{PhysRevLett.110.135704}%
  \BibitemOpen
  \bibfield  {author} {\bibinfo {author} {\bibfnamefont {M.}~\bibnamefont
  {Heyl}}, \bibinfo {author} {\bibfnamefont {A.}~\bibnamefont {Polkovnikov}},\
  and\ \bibinfo {author} {\bibfnamefont {S.}~\bibnamefont {Kehrein}},\
  }\bibfield  {title} {\bibinfo {title} {Dynamical quantum phase transitions in
  the transverse-field ising model},\ }\href
  {https://doi.org/10.1103/PhysRevLett.110.135704} {\bibfield  {journal}
  {\bibinfo  {journal} {Phys. Rev. Lett.}\ }\textbf {\bibinfo {volume} {110}},\
  \bibinfo {pages} {135704} (\bibinfo {year} {2013})}\BibitemShut {NoStop}%
\bibitem [{\citenamefont {Heyl}(2018)}]{Heyl_2018}%
  \BibitemOpen
  \bibfield  {author} {\bibinfo {author} {\bibfnamefont {M.}~\bibnamefont
  {Heyl}},\ }\bibfield  {title} {\bibinfo {title} {Dynamical quantum phase
  transitions: a review},\ }\href {https://doi.org/10.1088/1361-6633/aaaf9a}
  {\bibfield  {journal} {\bibinfo  {journal} {Reports on Progress in Physics}\
  }\textbf {\bibinfo {volume} {81}},\ \bibinfo {pages} {054001} (\bibinfo
  {year} {2018})}\BibitemShut {NoStop}%
\bibitem [{\citenamefont {Yang}\ and\ \citenamefont {Lee}(1952)}]{YangLee52i}%
  \BibitemOpen
  \bibfield  {author} {\bibinfo {author} {\bibfnamefont {C.~N.}\ \bibnamefont
  {Yang}}\ and\ \bibinfo {author} {\bibfnamefont {T.~D.}\ \bibnamefont {Lee}},\
  }\bibfield  {title} {\bibinfo {title} {Statistical theory of equations of
  state and phase transitions. i. theory of condensation},\ }\href
  {https://doi.org/10.1103/PhysRev.87.404} {\bibfield  {journal} {\bibinfo
  {journal} {Phys. Rev.}\ }\textbf {\bibinfo {volume} {87}},\ \bibinfo {pages}
  {404} (\bibinfo {year} {1952})}\BibitemShut {NoStop}%
\bibitem [{\citenamefont {Lee}\ and\ \citenamefont {Yang}(1952)}]{YangLee52ii}%
  \BibitemOpen
  \bibfield  {author} {\bibinfo {author} {\bibfnamefont {T.~D.}\ \bibnamefont
  {Lee}}\ and\ \bibinfo {author} {\bibfnamefont {C.~N.}\ \bibnamefont {Yang}},\
  }\bibfield  {title} {\bibinfo {title} {Statistical theory of equations of
  state and phase transitions. ii. lattice gas and ising model},\ }\href
  {https://doi.org/10.1103/PhysRev.87.410} {\bibfield  {journal} {\bibinfo
  {journal} {Phys. Rev.}\ }\textbf {\bibinfo {volume} {87}},\ \bibinfo {pages}
  {410} (\bibinfo {year} {1952})}\BibitemShut {NoStop}%
\bibitem [{\citenamefont {Fisher}\ and\ \citenamefont
  {Brittin}(1965)}]{fisher1965statistical}%
  \BibitemOpen
  \bibfield  {author} {\bibinfo {author} {\bibfnamefont {M.}~\bibnamefont
  {Fisher}}\ and\ \bibinfo {author} {\bibfnamefont {W.}~\bibnamefont
  {Brittin}},\ }\bibfield  {title} {\bibinfo {title} {Statistical physics, weak
  interactions, field theory},\ }\href@noop {} {\bibfield  {journal} {\bibinfo
  {journal} {Lectures in Theoretical Physics (Boulder: University of Colorado
  Press) vol VIIC}\ } (\bibinfo {year} {1965})}\BibitemShut {NoStop}%
\bibitem [{\citenamefont {Karrasch}\ and\ \citenamefont
  {Schuricht}(2013)}]{PhysRevB.87.195104}%
  \BibitemOpen
  \bibfield  {author} {\bibinfo {author} {\bibfnamefont {C.}~\bibnamefont
  {Karrasch}}\ and\ \bibinfo {author} {\bibfnamefont {D.}~\bibnamefont
  {Schuricht}},\ }\bibfield  {title} {\bibinfo {title} {Dynamical phase
  transitions after quenches in nonintegrable models},\ }\href
  {https://doi.org/10.1103/PhysRevB.87.195104} {\bibfield  {journal} {\bibinfo
  {journal} {Phys. Rev. B}\ }\textbf {\bibinfo {volume} {87}},\ \bibinfo
  {pages} {195104} (\bibinfo {year} {2013})}\BibitemShut {NoStop}%
\bibitem [{\citenamefont {Andraschko}\ and\ \citenamefont
  {Sirker}(2014)}]{PhysRevB.89.125120}%
  \BibitemOpen
  \bibfield  {author} {\bibinfo {author} {\bibfnamefont {F.}~\bibnamefont
  {Andraschko}}\ and\ \bibinfo {author} {\bibfnamefont {J.}~\bibnamefont
  {Sirker}},\ }\bibfield  {title} {\bibinfo {title} {Dynamical quantum phase
  transitions and the loschmidt echo: A transfer matrix approach},\ }\href
  {https://doi.org/10.1103/PhysRevB.89.125120} {\bibfield  {journal} {\bibinfo
  {journal} {Phys. Rev. B}\ }\textbf {\bibinfo {volume} {89}},\ \bibinfo
  {pages} {125120} (\bibinfo {year} {2014})}\BibitemShut {NoStop}%
\bibitem [{\citenamefont {Vajna}\ and\ \citenamefont
  {D\'ora}(2014)}]{PhysRevB.89.161105}%
  \BibitemOpen
  \bibfield  {author} {\bibinfo {author} {\bibfnamefont {S.}~\bibnamefont
  {Vajna}}\ and\ \bibinfo {author} {\bibfnamefont {B.}~\bibnamefont {D\'ora}},\
  }\bibfield  {title} {\bibinfo {title} {Disentangling dynamical phase
  transitions from equilibrium phase transitions},\ }\href
  {https://doi.org/10.1103/PhysRevB.89.161105} {\bibfield  {journal} {\bibinfo
  {journal} {Phys. Rev. B}\ }\textbf {\bibinfo {volume} {89}},\ \bibinfo
  {pages} {161105} (\bibinfo {year} {2014})}\BibitemShut {NoStop}%
\bibitem [{\citenamefont {Heyl}(2015)}]{PhysRevLett.115.140602}%
  \BibitemOpen
  \bibfield  {author} {\bibinfo {author} {\bibfnamefont {M.}~\bibnamefont
  {Heyl}},\ }\bibfield  {title} {\bibinfo {title} {Scaling and universality at
  dynamical quantum phase transitions},\ }\href
  {https://doi.org/10.1103/PhysRevLett.115.140602} {\bibfield  {journal}
  {\bibinfo  {journal} {Phys. Rev. Lett.}\ }\textbf {\bibinfo {volume} {115}},\
  \bibinfo {pages} {140602} (\bibinfo {year} {2015})}\BibitemShut {NoStop}%
\bibitem [{\citenamefont {Abeling}\ and\ \citenamefont
  {Kehrein}(2016)}]{PhysRevB.93.104302}%
  \BibitemOpen
  \bibfield  {author} {\bibinfo {author} {\bibfnamefont {N.~O.}\ \bibnamefont
  {Abeling}}\ and\ \bibinfo {author} {\bibfnamefont {S.}~\bibnamefont
  {Kehrein}},\ }\bibfield  {title} {\bibinfo {title} {Quantum quench dynamics
  in the transverse field ising model at nonzero temperatures},\ }\href
  {https://doi.org/10.1103/PhysRevB.93.104302} {\bibfield  {journal} {\bibinfo
  {journal} {Phys. Rev. B}\ }\textbf {\bibinfo {volume} {93}},\ \bibinfo
  {pages} {104302} (\bibinfo {year} {2016})}\BibitemShut {NoStop}%
\bibitem [{\citenamefont {Sharma}\ \emph {et~al.}(2016)\citenamefont {Sharma},
  \citenamefont {Divakaran}, \citenamefont {Polkovnikov},\ and\ \citenamefont
  {Dutta}}]{PhysRevB.93.144306}%
  \BibitemOpen
  \bibfield  {author} {\bibinfo {author} {\bibfnamefont {S.}~\bibnamefont
  {Sharma}}, \bibinfo {author} {\bibfnamefont {U.}~\bibnamefont {Divakaran}},
  \bibinfo {author} {\bibfnamefont {A.}~\bibnamefont {Polkovnikov}},\ and\
  \bibinfo {author} {\bibfnamefont {A.}~\bibnamefont {Dutta}},\ }\bibfield
  {title} {\bibinfo {title} {Slow quenches in a quantum ising chain: Dynamical
  phase transitions and topology},\ }\href
  {https://doi.org/10.1103/PhysRevB.93.144306} {\bibfield  {journal} {\bibinfo
  {journal} {Phys. Rev. B}\ }\textbf {\bibinfo {volume} {93}},\ \bibinfo
  {pages} {144306} (\bibinfo {year} {2016})}\BibitemShut {NoStop}%
\bibitem [{\citenamefont {Halimeh}\ and\ \citenamefont
  {Zauner-Stauber}(2017)}]{PhysRevB.96.134427}%
  \BibitemOpen
  \bibfield  {author} {\bibinfo {author} {\bibfnamefont {J.~C.}\ \bibnamefont
  {Halimeh}}\ and\ \bibinfo {author} {\bibfnamefont {V.}~\bibnamefont
  {Zauner-Stauber}},\ }\bibfield  {title} {\bibinfo {title} {Dynamical phase
  diagram of quantum spin chains with long-range interactions},\ }\href
  {https://doi.org/10.1103/PhysRevB.96.134427} {\bibfield  {journal} {\bibinfo
  {journal} {Phys. Rev. B}\ }\textbf {\bibinfo {volume} {96}},\ \bibinfo
  {pages} {134427} (\bibinfo {year} {2017})}\BibitemShut {NoStop}%
\bibitem [{\citenamefont {\ifmmode \check{Z}\else
  \v{Z}\fi{}unkovi\ifmmode~\check{c}\else \v{c}\fi{}}\ \emph
  {et~al.}(2018)\citenamefont {\ifmmode \check{Z}\else
  \v{Z}\fi{}unkovi\ifmmode~\check{c}\else \v{c}\fi{}}, \citenamefont {Heyl},
  \citenamefont {Knap},\ and\ \citenamefont {Silva}}]{PhysRevLett.120.130601}%
  \BibitemOpen
  \bibfield  {author} {\bibinfo {author} {\bibfnamefont {B.}~\bibnamefont
  {\ifmmode \check{Z}\else \v{Z}\fi{}unkovi\ifmmode~\check{c}\else
  \v{c}\fi{}}}, \bibinfo {author} {\bibfnamefont {M.}~\bibnamefont {Heyl}},
  \bibinfo {author} {\bibfnamefont {M.}~\bibnamefont {Knap}},\ and\ \bibinfo
  {author} {\bibfnamefont {A.}~\bibnamefont {Silva}},\ }\bibfield  {title}
  {\bibinfo {title} {Dynamical quantum phase transitions in spin chains with
  long-range interactions: Merging different concepts of nonequilibrium
  criticality},\ }\href {https://doi.org/10.1103/PhysRevLett.120.130601}
  {\bibfield  {journal} {\bibinfo  {journal} {Phys. Rev. Lett.}\ }\textbf
  {\bibinfo {volume} {120}},\ \bibinfo {pages} {130601} (\bibinfo {year}
  {2018})}\BibitemShut {NoStop}%
\bibitem [{\citenamefont {Hamazaki}(2021)}]{Hamazaki2021}%
  \BibitemOpen
  \bibfield  {author} {\bibinfo {author} {\bibfnamefont {R.}~\bibnamefont
  {Hamazaki}},\ }\bibfield  {title} {\bibinfo {title} {Exceptional dynamical
  quantum phase transitions in periodically driven systems},\ }\href
  {https://doi.org/10.1038/s41467-021-25355-3} {\bibfield  {journal} {\bibinfo
  {journal} {Nature Communications}\ }\textbf {\bibinfo {volume} {12}},\
  \bibinfo {pages} {5108} (\bibinfo {year} {2021})}\BibitemShut {NoStop}%
\bibitem [{\citenamefont {Halimeh}\ \emph {et~al.}(2021)\citenamefont
  {Halimeh}, \citenamefont {Trapin}, \citenamefont {Van~Damme},\ and\
  \citenamefont {Heyl}}]{PhysRevB.104.075130}%
  \BibitemOpen
  \bibfield  {author} {\bibinfo {author} {\bibfnamefont {J.~C.}\ \bibnamefont
  {Halimeh}}, \bibinfo {author} {\bibfnamefont {D.}~\bibnamefont {Trapin}},
  \bibinfo {author} {\bibfnamefont {M.}~\bibnamefont {Van~Damme}},\ and\
  \bibinfo {author} {\bibfnamefont {M.}~\bibnamefont {Heyl}},\ }\bibfield
  {title} {\bibinfo {title} {Local measures of dynamical quantum phase
  transitions},\ }\href {https://doi.org/10.1103/PhysRevB.104.075130}
  {\bibfield  {journal} {\bibinfo  {journal} {Phys. Rev. B}\ }\textbf {\bibinfo
  {volume} {104}},\ \bibinfo {pages} {075130} (\bibinfo {year}
  {2021})}\BibitemShut {NoStop}%
\bibitem [{\citenamefont {De~Nicola}\ \emph {et~al.}(2021)\citenamefont
  {De~Nicola}, \citenamefont {Michailidis},\ and\ \citenamefont
  {Serbyn}}]{PhysRevLett.126.040602}%
  \BibitemOpen
  \bibfield  {author} {\bibinfo {author} {\bibfnamefont {S.}~\bibnamefont
  {De~Nicola}}, \bibinfo {author} {\bibfnamefont {A.~A.}\ \bibnamefont
  {Michailidis}},\ and\ \bibinfo {author} {\bibfnamefont {M.}~\bibnamefont
  {Serbyn}},\ }\bibfield  {title} {\bibinfo {title} {Entanglement view of
  dynamical quantum phase transitions},\ }\href
  {https://doi.org/10.1103/PhysRevLett.126.040602} {\bibfield  {journal}
  {\bibinfo  {journal} {Phys. Rev. Lett.}\ }\textbf {\bibinfo {volume} {126}},\
  \bibinfo {pages} {040602} (\bibinfo {year} {2021})}\BibitemShut {NoStop}%
\bibitem [{\citenamefont {Hashizume}\ \emph {et~al.}(2022)\citenamefont
  {Hashizume}, \citenamefont {McCulloch},\ and\ \citenamefont
  {Halimeh}}]{PhysRevResearch.4.013250}%
  \BibitemOpen
  \bibfield  {author} {\bibinfo {author} {\bibfnamefont {T.}~\bibnamefont
  {Hashizume}}, \bibinfo {author} {\bibfnamefont {I.~P.}\ \bibnamefont
  {McCulloch}},\ and\ \bibinfo {author} {\bibfnamefont {J.~C.}\ \bibnamefont
  {Halimeh}},\ }\bibfield  {title} {\bibinfo {title} {Dynamical phase
  transitions in the two-dimensional transverse-field ising model},\ }\href
  {https://doi.org/10.1103/PhysRevResearch.4.013250} {\bibfield  {journal}
  {\bibinfo  {journal} {Phys. Rev. Res.}\ }\textbf {\bibinfo {volume} {4}},\
  \bibinfo {pages} {013250} (\bibinfo {year} {2022})}\BibitemShut {NoStop}%
\bibitem [{\citenamefont {Osborne}\ \emph {et~al.}(2025)\citenamefont
  {Osborne}, \citenamefont {McCulloch},\ and\ \citenamefont
  {Halimeh}}]{rnv5-f32k}%
  \BibitemOpen
  \bibfield  {author} {\bibinfo {author} {\bibfnamefont {J.~J.}\ \bibnamefont
  {Osborne}}, \bibinfo {author} {\bibfnamefont {I.~P.}\ \bibnamefont
  {McCulloch}},\ and\ \bibinfo {author} {\bibfnamefont {J.~C.}\ \bibnamefont
  {Halimeh}},\ }\bibfield  {title} {\bibinfo {title} {Probing confinement
  through dynamical quantum phase transitions: From quantum spin models to
  lattice gauge theories},\ }\href {https://doi.org/10.1103/rnv5-f32k}
  {\bibfield  {journal} {\bibinfo  {journal} {Phys. Rev. Res.}\ }\textbf
  {\bibinfo {volume} {7}},\ \bibinfo {pages} {043076} (\bibinfo {year}
  {2025})}\BibitemShut {NoStop}%
\bibitem [{\citenamefont {Khatun}\ and\ \citenamefont
  {Bhattacharjee}(2019)}]{PhysRevLett.123.160603}%
  \BibitemOpen
  \bibfield  {author} {\bibinfo {author} {\bibfnamefont {A.}~\bibnamefont
  {Khatun}}\ and\ \bibinfo {author} {\bibfnamefont {S.~M.}\ \bibnamefont
  {Bhattacharjee}},\ }\bibfield  {title} {\bibinfo {title} {Boundaries and
  unphysical fixed points in dynamical quantum phase transitions},\ }\href
  {https://doi.org/10.1103/PhysRevLett.123.160603} {\bibfield  {journal}
  {\bibinfo  {journal} {Phys. Rev. Lett.}\ }\textbf {\bibinfo {volume} {123}},\
  \bibinfo {pages} {160603} (\bibinfo {year} {2019})}\BibitemShut {NoStop}%
\bibitem [{\citenamefont {Jurcevic}\ \emph {et~al.}(2017)\citenamefont
  {Jurcevic}, \citenamefont {Shen}, \citenamefont {Hauke}, \citenamefont
  {Maier}, \citenamefont {Brydges}, \citenamefont {Hempel}, \citenamefont
  {Lanyon}, \citenamefont {Heyl}, \citenamefont {Blatt},\ and\ \citenamefont
  {Roos}}]{PhysRevLett.119.080501}%
  \BibitemOpen
  \bibfield  {author} {\bibinfo {author} {\bibfnamefont {P.}~\bibnamefont
  {Jurcevic}}, \bibinfo {author} {\bibfnamefont {H.}~\bibnamefont {Shen}},
  \bibinfo {author} {\bibfnamefont {P.}~\bibnamefont {Hauke}}, \bibinfo
  {author} {\bibfnamefont {C.}~\bibnamefont {Maier}}, \bibinfo {author}
  {\bibfnamefont {T.}~\bibnamefont {Brydges}}, \bibinfo {author} {\bibfnamefont
  {C.}~\bibnamefont {Hempel}}, \bibinfo {author} {\bibfnamefont {B.~P.}\
  \bibnamefont {Lanyon}}, \bibinfo {author} {\bibfnamefont {M.}~\bibnamefont
  {Heyl}}, \bibinfo {author} {\bibfnamefont {R.}~\bibnamefont {Blatt}},\ and\
  \bibinfo {author} {\bibfnamefont {C.~F.}\ \bibnamefont {Roos}},\ }\bibfield
  {title} {\bibinfo {title} {Direct observation of dynamical quantum phase
  transitions in an interacting many-body system},\ }\href
  {https://doi.org/10.1103/PhysRevLett.119.080501} {\bibfield  {journal}
  {\bibinfo  {journal} {Phys. Rev. Lett.}\ }\textbf {\bibinfo {volume} {119}},\
  \bibinfo {pages} {080501} (\bibinfo {year} {2017})}\BibitemShut {NoStop}%
\bibitem [{\citenamefont {Fl{\"a}schner}\ \emph {et~al.}(2018)\citenamefont
  {Fl{\"a}schner}, \citenamefont {Vogel}, \citenamefont {Tarnowski},
  \citenamefont {Rem}, \citenamefont {L{\"u}hmann}, \citenamefont {Heyl},
  \citenamefont {Budich}, \citenamefont {Mathey}, \citenamefont {Sengstock},\
  and\ \citenamefont {Weitenberg}}]{Flaschner2018}%
  \BibitemOpen
  \bibfield  {author} {\bibinfo {author} {\bibfnamefont {N.}~\bibnamefont
  {Fl{\"a}schner}}, \bibinfo {author} {\bibfnamefont {D.}~\bibnamefont
  {Vogel}}, \bibinfo {author} {\bibfnamefont {M.}~\bibnamefont {Tarnowski}},
  \bibinfo {author} {\bibfnamefont {B.~S.}\ \bibnamefont {Rem}}, \bibinfo
  {author} {\bibfnamefont {D.-S.}\ \bibnamefont {L{\"u}hmann}}, \bibinfo
  {author} {\bibfnamefont {M.}~\bibnamefont {Heyl}}, \bibinfo {author}
  {\bibfnamefont {J.~C.}\ \bibnamefont {Budich}}, \bibinfo {author}
  {\bibfnamefont {L.}~\bibnamefont {Mathey}}, \bibinfo {author} {\bibfnamefont
  {K.}~\bibnamefont {Sengstock}},\ and\ \bibinfo {author} {\bibfnamefont
  {C.}~\bibnamefont {Weitenberg}},\ }\bibfield  {title} {\bibinfo {title}
  {Observation of dynamical vortices after quenches in a system with
  topology},\ }\href {https://doi.org/10.1038/s41567-017-0013-8} {\bibfield
  {journal} {\bibinfo  {journal} {Nature Physics}\ }\textbf {\bibinfo {volume}
  {14}},\ \bibinfo {pages} {265} (\bibinfo {year} {2018})}\BibitemShut
  {NoStop}%
\bibitem [{\citenamefont {Guo}\ \emph {et~al.}(2019)\citenamefont {Guo},
  \citenamefont {Yang}, \citenamefont {Zeng}, \citenamefont {Peng},
  \citenamefont {Li}, \citenamefont {Deng}, \citenamefont {Jin}, \citenamefont
  {Chen}, \citenamefont {Zheng},\ and\ \citenamefont
  {Fan}}]{PhysRevApplied.11.044080}%
  \BibitemOpen
  \bibfield  {author} {\bibinfo {author} {\bibfnamefont {X.-Y.}\ \bibnamefont
  {Guo}}, \bibinfo {author} {\bibfnamefont {C.}~\bibnamefont {Yang}}, \bibinfo
  {author} {\bibfnamefont {Y.}~\bibnamefont {Zeng}}, \bibinfo {author}
  {\bibfnamefont {Y.}~\bibnamefont {Peng}}, \bibinfo {author} {\bibfnamefont
  {H.-K.}\ \bibnamefont {Li}}, \bibinfo {author} {\bibfnamefont
  {H.}~\bibnamefont {Deng}}, \bibinfo {author} {\bibfnamefont {Y.-R.}\
  \bibnamefont {Jin}}, \bibinfo {author} {\bibfnamefont {S.}~\bibnamefont
  {Chen}}, \bibinfo {author} {\bibfnamefont {D.}~\bibnamefont {Zheng}},\ and\
  \bibinfo {author} {\bibfnamefont {H.}~\bibnamefont {Fan}},\ }\bibfield
  {title} {\bibinfo {title} {Observation of a dynamical quantum phase
  transition by a superconducting qubit simulation},\ }\href
  {https://doi.org/10.1103/PhysRevApplied.11.044080} {\bibfield  {journal}
  {\bibinfo  {journal} {Phys. Rev. Appl.}\ }\textbf {\bibinfo {volume} {11}},\
  \bibinfo {pages} {044080} (\bibinfo {year} {2019})}\BibitemShut {NoStop}%
\bibitem [{\citenamefont {Chen}\ \emph {et~al.}(2020)\citenamefont {Chen},
  \citenamefont {Cui}, \citenamefont {Ai}, \citenamefont {He}, \citenamefont
  {Huang}, \citenamefont {Han}, \citenamefont {Li},\ and\ \citenamefont
  {Guo}}]{PhysRevA.102.042222}%
  \BibitemOpen
  \bibfield  {author} {\bibinfo {author} {\bibfnamefont {Z.}~\bibnamefont
  {Chen}}, \bibinfo {author} {\bibfnamefont {J.-M.}\ \bibnamefont {Cui}},
  \bibinfo {author} {\bibfnamefont {M.-Z.}\ \bibnamefont {Ai}}, \bibinfo
  {author} {\bibfnamefont {R.}~\bibnamefont {He}}, \bibinfo {author}
  {\bibfnamefont {Y.-F.}\ \bibnamefont {Huang}}, \bibinfo {author}
  {\bibfnamefont {Y.-J.}\ \bibnamefont {Han}}, \bibinfo {author} {\bibfnamefont
  {C.-F.}\ \bibnamefont {Li}},\ and\ \bibinfo {author} {\bibfnamefont {G.-C.}\
  \bibnamefont {Guo}},\ }\bibfield  {title} {\bibinfo {title} {Experimentally
  detecting dynamical quantum phase transitions in a slowly quenched
  ising-chain model},\ }\href {https://doi.org/10.1103/PhysRevA.102.042222}
  {\bibfield  {journal} {\bibinfo  {journal} {Phys. Rev. A}\ }\textbf {\bibinfo
  {volume} {102}},\ \bibinfo {pages} {042222} (\bibinfo {year}
  {2020})}\BibitemShut {NoStop}%
\bibitem [{\citenamefont {Xu}\ \emph {et~al.}(2020)\citenamefont {Xu},
  \citenamefont {Sun}, \citenamefont {Liu}, \citenamefont {Zhang},
  \citenamefont {Li}, \citenamefont {Dong}, \citenamefont {Ren}, \citenamefont
  {Zhang}, \citenamefont {Nori}, \citenamefont {Zheng}, \citenamefont {Fan},\
  and\ \citenamefont {Wang}}]{sciadv.aba4935}%
  \BibitemOpen
  \bibfield  {author} {\bibinfo {author} {\bibfnamefont {K.}~\bibnamefont
  {Xu}}, \bibinfo {author} {\bibfnamefont {Z.-H.}\ \bibnamefont {Sun}},
  \bibinfo {author} {\bibfnamefont {W.}~\bibnamefont {Liu}}, \bibinfo {author}
  {\bibfnamefont {Y.-R.}\ \bibnamefont {Zhang}}, \bibinfo {author}
  {\bibfnamefont {H.}~\bibnamefont {Li}}, \bibinfo {author} {\bibfnamefont
  {H.}~\bibnamefont {Dong}}, \bibinfo {author} {\bibfnamefont {W.}~\bibnamefont
  {Ren}}, \bibinfo {author} {\bibfnamefont {P.}~\bibnamefont {Zhang}}, \bibinfo
  {author} {\bibfnamefont {F.}~\bibnamefont {Nori}}, \bibinfo {author}
  {\bibfnamefont {D.}~\bibnamefont {Zheng}}, \bibinfo {author} {\bibfnamefont
  {H.}~\bibnamefont {Fan}},\ and\ \bibinfo {author} {\bibfnamefont
  {H.}~\bibnamefont {Wang}},\ }\bibfield  {title} {\bibinfo {title} {Probing
  dynamical phase transitions with a superconducting quantum simulator},\
  }\href {https://doi.org/10.1126/sciadv.aba4935} {\bibfield  {journal}
  {\bibinfo  {journal} {Science Advances}\ }\textbf {\bibinfo {volume} {6}},\
  \bibinfo {pages} {eaba4935} (\bibinfo {year} {2020})},\ \Eprint
  {https://arxiv.org/abs/https://www.science.org/doi/pdf/10.1126/sciadv.aba4935}
  {https://www.science.org/doi/pdf/10.1126/sciadv.aba4935} \BibitemShut
  {NoStop}%
\bibitem [{\citenamefont {Lund}\ \emph {et~al.}(2017)\citenamefont {Lund},
  \citenamefont {Bremner},\ and\ \citenamefont {Ralph}}]{Lund2017}%
  \BibitemOpen
  \bibfield  {author} {\bibinfo {author} {\bibfnamefont {A.~P.}\ \bibnamefont
  {Lund}}, \bibinfo {author} {\bibfnamefont {M.~J.}\ \bibnamefont {Bremner}},\
  and\ \bibinfo {author} {\bibfnamefont {T.~C.}\ \bibnamefont {Ralph}},\
  }\bibfield  {title} {\bibinfo {title} {Quantum sampling problems,
  bosonsampling and quantum supremacy},\ }\href
  {https://doi.org/10.1038/s41534-017-0018-2} {\bibfield  {journal} {\bibinfo
  {journal} {npj Quantum Information}\ }\textbf {\bibinfo {volume} {3}},\
  \bibinfo {pages} {15} (\bibinfo {year} {2017})}\BibitemShut {NoStop}%
\bibitem [{\citenamefont {Bouland}\ \emph {et~al.}(2019)\citenamefont
  {Bouland}, \citenamefont {Fefferman}, \citenamefont {Nirkhe},\ and\
  \citenamefont {Vazirani}}]{Bouland2019}%
  \BibitemOpen
  \bibfield  {author} {\bibinfo {author} {\bibfnamefont {A.}~\bibnamefont
  {Bouland}}, \bibinfo {author} {\bibfnamefont {B.}~\bibnamefont {Fefferman}},
  \bibinfo {author} {\bibfnamefont {C.}~\bibnamefont {Nirkhe}},\ and\ \bibinfo
  {author} {\bibfnamefont {U.}~\bibnamefont {Vazirani}},\ }\bibfield  {title}
  {\bibinfo {title} {On the complexity and verification of quantum random
  circuit sampling},\ }\href {https://doi.org/10.1038/s41567-018-0318-2}
  {\bibfield  {journal} {\bibinfo  {journal} {Nature Physics}\ }\textbf
  {\bibinfo {volume} {15}},\ \bibinfo {pages} {159} (\bibinfo {year}
  {2019})}\BibitemShut {NoStop}%
\bibitem [{\citenamefont {Arute}\ \emph {et~al.}(2019)\citenamefont {Arute},
  \citenamefont {Arya}, \citenamefont {Babbush}, \citenamefont {Bacon},
  \citenamefont {Bardin}, \citenamefont {Barends}, \citenamefont {Biswas},
  \citenamefont {Boixo}, \citenamefont {Brandao}, \citenamefont {Buell},
  \citenamefont {Burkett}, \citenamefont {Chen}, \citenamefont {Chen},
  \citenamefont {Chiaro}, \citenamefont {Collins}, \citenamefont {Courtney},
  \citenamefont {Dunsworth}, \citenamefont {Farhi}, \citenamefont {Foxen},
  \citenamefont {Fowler}, \citenamefont {Gidney}, \citenamefont {Giustina},
  \citenamefont {Graff}, \citenamefont {Guerin}, \citenamefont {Habegger},
  \citenamefont {Harrigan}, \citenamefont {Hartmann}, \citenamefont {Ho},
  \citenamefont {Hoffmann}, \citenamefont {Huang}, \citenamefont {Humble},
  \citenamefont {Isakov}, \citenamefont {Jeffrey}, \citenamefont {Jiang},
  \citenamefont {Kafri}, \citenamefont {Kechedzhi}, \citenamefont {Kelly},
  \citenamefont {Klimov}, \citenamefont {Knysh}, \citenamefont {Korotkov},
  \citenamefont {Kostritsa}, \citenamefont {Landhuis}, \citenamefont
  {Lindmark}, \citenamefont {Lucero}, \citenamefont {Lyakh}, \citenamefont
  {Mandrà}, \citenamefont {McClean}, \citenamefont {McEwen}, \citenamefont
  {Megrant}, \citenamefont {Mi}, \citenamefont {Michielsen}, \citenamefont
  {Mohseni}, \citenamefont {Mutus}, \citenamefont {Naaman}, \citenamefont
  {Neeley}, \citenamefont {Neill}, \citenamefont {Niu}, \citenamefont {Ostby},
  \citenamefont {Petukhov}, \citenamefont {Platt}, \citenamefont {Quintana},
  \citenamefont {Rieffel}, \citenamefont {Roushan}, \citenamefont {Rubin},
  \citenamefont {Sank}, \citenamefont {Satzinger}, \citenamefont {Smelyanskiy},
  \citenamefont {Sung}, \citenamefont {Trevithick}, \citenamefont
  {Vainsencher}, \citenamefont {Villalonga}, \citenamefont {White},
  \citenamefont {Yao}, \citenamefont {Yeh}, \citenamefont {Zalcman},
  \citenamefont {Neven},\ and\ \citenamefont {Martinis}}]{Arute2019}%
  \BibitemOpen
  \bibfield  {author} {\bibinfo {author} {\bibfnamefont {F.}~\bibnamefont
  {Arute}}, \bibinfo {author} {\bibfnamefont {K.}~\bibnamefont {Arya}},
  \bibinfo {author} {\bibfnamefont {R.}~\bibnamefont {Babbush}}, \bibinfo
  {author} {\bibfnamefont {D.}~\bibnamefont {Bacon}}, \bibinfo {author}
  {\bibfnamefont {J.~C.}\ \bibnamefont {Bardin}}, \bibinfo {author}
  {\bibfnamefont {R.}~\bibnamefont {Barends}}, \bibinfo {author} {\bibfnamefont
  {R.}~\bibnamefont {Biswas}}, \bibinfo {author} {\bibfnamefont
  {S.}~\bibnamefont {Boixo}}, \bibinfo {author} {\bibfnamefont {F.~G. S.~L.}\
  \bibnamefont {Brandao}}, \bibinfo {author} {\bibfnamefont {D.~A.}\
  \bibnamefont {Buell}}, \bibinfo {author} {\bibfnamefont {B.}~\bibnamefont
  {Burkett}}, \bibinfo {author} {\bibfnamefont {Y.}~\bibnamefont {Chen}},
  \bibinfo {author} {\bibfnamefont {Z.}~\bibnamefont {Chen}}, \bibinfo {author}
  {\bibfnamefont {B.}~\bibnamefont {Chiaro}}, \bibinfo {author} {\bibfnamefont
  {R.}~\bibnamefont {Collins}}, \bibinfo {author} {\bibfnamefont
  {W.}~\bibnamefont {Courtney}}, \bibinfo {author} {\bibfnamefont
  {A.}~\bibnamefont {Dunsworth}}, \bibinfo {author} {\bibfnamefont
  {E.}~\bibnamefont {Farhi}}, \bibinfo {author} {\bibfnamefont
  {B.}~\bibnamefont {Foxen}}, \bibinfo {author} {\bibfnamefont
  {A.}~\bibnamefont {Fowler}}, \bibinfo {author} {\bibfnamefont
  {C.}~\bibnamefont {Gidney}}, \bibinfo {author} {\bibfnamefont
  {M.}~\bibnamefont {Giustina}}, \bibinfo {author} {\bibfnamefont
  {R.}~\bibnamefont {Graff}}, \bibinfo {author} {\bibfnamefont
  {K.}~\bibnamefont {Guerin}}, \bibinfo {author} {\bibfnamefont
  {S.}~\bibnamefont {Habegger}}, \bibinfo {author} {\bibfnamefont {M.~P.}\
  \bibnamefont {Harrigan}}, \bibinfo {author} {\bibfnamefont {M.~J.}\
  \bibnamefont {Hartmann}}, \bibinfo {author} {\bibfnamefont {A.}~\bibnamefont
  {Ho}}, \bibinfo {author} {\bibfnamefont {M.}~\bibnamefont {Hoffmann}},
  \bibinfo {author} {\bibfnamefont {T.}~\bibnamefont {Huang}}, \bibinfo
  {author} {\bibfnamefont {T.~S.}\ \bibnamefont {Humble}}, \bibinfo {author}
  {\bibfnamefont {S.~V.}\ \bibnamefont {Isakov}}, \bibinfo {author}
  {\bibfnamefont {E.}~\bibnamefont {Jeffrey}}, \bibinfo {author} {\bibfnamefont
  {Z.}~\bibnamefont {Jiang}}, \bibinfo {author} {\bibfnamefont
  {D.}~\bibnamefont {Kafri}}, \bibinfo {author} {\bibfnamefont
  {K.}~\bibnamefont {Kechedzhi}}, \bibinfo {author} {\bibfnamefont
  {J.}~\bibnamefont {Kelly}}, \bibinfo {author} {\bibfnamefont {P.~V.}\
  \bibnamefont {Klimov}}, \bibinfo {author} {\bibfnamefont {S.}~\bibnamefont
  {Knysh}}, \bibinfo {author} {\bibfnamefont {A.}~\bibnamefont {Korotkov}},
  \bibinfo {author} {\bibfnamefont {F.}~\bibnamefont {Kostritsa}}, \bibinfo
  {author} {\bibfnamefont {D.}~\bibnamefont {Landhuis}}, \bibinfo {author}
  {\bibfnamefont {M.}~\bibnamefont {Lindmark}}, \bibinfo {author}
  {\bibfnamefont {E.}~\bibnamefont {Lucero}}, \bibinfo {author} {\bibfnamefont
  {D.}~\bibnamefont {Lyakh}}, \bibinfo {author} {\bibfnamefont
  {S.}~\bibnamefont {Mandrà}}, \bibinfo {author} {\bibfnamefont {J.~R.}\
  \bibnamefont {McClean}}, \bibinfo {author} {\bibfnamefont {M.}~\bibnamefont
  {McEwen}}, \bibinfo {author} {\bibfnamefont {A.}~\bibnamefont {Megrant}},
  \bibinfo {author} {\bibfnamefont {X.}~\bibnamefont {Mi}}, \bibinfo {author}
  {\bibfnamefont {K.}~\bibnamefont {Michielsen}}, \bibinfo {author}
  {\bibfnamefont {M.}~\bibnamefont {Mohseni}}, \bibinfo {author} {\bibfnamefont
  {J.}~\bibnamefont {Mutus}}, \bibinfo {author} {\bibfnamefont
  {O.}~\bibnamefont {Naaman}}, \bibinfo {author} {\bibfnamefont
  {M.}~\bibnamefont {Neeley}}, \bibinfo {author} {\bibfnamefont
  {C.}~\bibnamefont {Neill}}, \bibinfo {author} {\bibfnamefont {M.~Y.}\
  \bibnamefont {Niu}}, \bibinfo {author} {\bibfnamefont {E.}~\bibnamefont
  {Ostby}}, \bibinfo {author} {\bibfnamefont {A.}~\bibnamefont {Petukhov}},
  \bibinfo {author} {\bibfnamefont {J.~C.}\ \bibnamefont {Platt}}, \bibinfo
  {author} {\bibfnamefont {C.}~\bibnamefont {Quintana}}, \bibinfo {author}
  {\bibfnamefont {E.~G.}\ \bibnamefont {Rieffel}}, \bibinfo {author}
  {\bibfnamefont {P.}~\bibnamefont {Roushan}}, \bibinfo {author} {\bibfnamefont
  {N.~C.}\ \bibnamefont {Rubin}}, \bibinfo {author} {\bibfnamefont
  {D.}~\bibnamefont {Sank}}, \bibinfo {author} {\bibfnamefont {K.~J.}\
  \bibnamefont {Satzinger}}, \bibinfo {author} {\bibfnamefont {V.}~\bibnamefont
  {Smelyanskiy}}, \bibinfo {author} {\bibfnamefont {K.~J.}\ \bibnamefont
  {Sung}}, \bibinfo {author} {\bibfnamefont {M.~D.}\ \bibnamefont
  {Trevithick}}, \bibinfo {author} {\bibfnamefont {A.}~\bibnamefont
  {Vainsencher}}, \bibinfo {author} {\bibfnamefont {B.}~\bibnamefont
  {Villalonga}}, \bibinfo {author} {\bibfnamefont {T.}~\bibnamefont {White}},
  \bibinfo {author} {\bibfnamefont {Z.~J.}\ \bibnamefont {Yao}}, \bibinfo
  {author} {\bibfnamefont {P.}~\bibnamefont {Yeh}}, \bibinfo {author}
  {\bibfnamefont {A.}~\bibnamefont {Zalcman}}, \bibinfo {author} {\bibfnamefont
  {H.}~\bibnamefont {Neven}},\ and\ \bibinfo {author} {\bibfnamefont {J.~M.}\
  \bibnamefont {Martinis}},\ }\bibfield  {title} {\bibinfo {title} {Quantum
  supremacy using a programmable superconducting processor},\ }\href
  {https://doi.org/10.1038/s41586-019-1666-5} {\bibfield  {journal} {\bibinfo
  {journal} {Nature}\ }\textbf {\bibinfo {volume} {574}},\ \bibinfo {pages}
  {505} (\bibinfo {year} {2019})}\BibitemShut {NoStop}%
\bibitem [{\citenamefont {Lee}\ \emph {et~al.}(2023)\citenamefont {Lee},
  \citenamefont {Jian},\ and\ \citenamefont {Xu}}]{LeeJianXu}%
  \BibitemOpen
  \bibfield  {author} {\bibinfo {author} {\bibfnamefont {J.~Y.}\ \bibnamefont
  {Lee}}, \bibinfo {author} {\bibfnamefont {C.-M.}\ \bibnamefont {Jian}},\ and\
  \bibinfo {author} {\bibfnamefont {C.}~\bibnamefont {Xu}},\ }\bibfield
  {title} {\bibinfo {title} {{Quantum Criticality Under Decoherence or Weak
  Measurement}},\ }\href {https://doi.org/10.1103/PRXQuantum.4.030317}
  {\bibfield  {journal} {\bibinfo  {journal} {PRX Quantum}\ }\textbf {\bibinfo
  {volume} {4}},\ \bibinfo {pages} {030317} (\bibinfo {year}
  {2023})}\BibitemShut {NoStop}%
\bibitem [{\citenamefont {Fan}\ \emph {et~al.}(2024)\citenamefont {Fan},
  \citenamefont {Bao}, \citenamefont {Altman},\ and\ \citenamefont
  {Vishwanath}}]{FanBaoAltmanVishwanath}%
  \BibitemOpen
  \bibfield  {author} {\bibinfo {author} {\bibfnamefont {R.}~\bibnamefont
  {Fan}}, \bibinfo {author} {\bibfnamefont {Y.}~\bibnamefont {Bao}}, \bibinfo
  {author} {\bibfnamefont {E.}~\bibnamefont {Altman}},\ and\ \bibinfo {author}
  {\bibfnamefont {A.}~\bibnamefont {Vishwanath}},\ }\bibfield  {title}
  {\bibinfo {title} {{Diagnostics of Mixed-State Topological Order and
  Breakdown of Quantum Memory}},\ }\href
  {https://doi.org/10.1103/PRXQuantum.5.020343} {\bibfield  {journal} {\bibinfo
   {journal} {PRX Quantum}\ }\textbf {\bibinfo {volume} {5}},\ \bibinfo {pages}
  {020343} (\bibinfo {year} {2024})}\BibitemShut {NoStop}%
\bibitem [{\citenamefont {Sala}\ \emph
  {et~al.}(2024{\natexlab{a}})\citenamefont {Sala}, \citenamefont
  {Gopalakrishnan}, \citenamefont {Oshikawa},\ and\ \citenamefont
  {You}}]{SalaGopalakrishnanOshikawaYou}%
  \BibitemOpen
  \bibfield  {author} {\bibinfo {author} {\bibfnamefont {P.}~\bibnamefont
  {Sala}}, \bibinfo {author} {\bibfnamefont {S.}~\bibnamefont
  {Gopalakrishnan}}, \bibinfo {author} {\bibfnamefont {M.}~\bibnamefont
  {Oshikawa}},\ and\ \bibinfo {author} {\bibfnamefont {Y.}~\bibnamefont
  {You}},\ }\bibfield  {title} {\bibinfo {title} {{Spontaneous strong symmetry
  breaking in open systems: Purification perspective}},\ }\href
  {https://doi.org/10.1103/PhysRevB.110.155150} {\bibfield  {journal} {\bibinfo
   {journal} {Phys. Rev. B}\ }\textbf {\bibinfo {volume} {110}},\ \bibinfo
  {pages} {155150} (\bibinfo {year} {2024}{\natexlab{a}})}\BibitemShut
  {NoStop}%
\bibitem [{\citenamefont {Ellison}\ and\ \citenamefont
  {Cheng}(2025)}]{EllisonCheng}%
  \BibitemOpen
  \bibfield  {author} {\bibinfo {author} {\bibfnamefont {T.~D.}\ \bibnamefont
  {Ellison}}\ and\ \bibinfo {author} {\bibfnamefont {M.}~\bibnamefont
  {Cheng}},\ }\bibfield  {title} {\bibinfo {title} {{Toward a Classification of
  Mixed-State Topological Orders in Two Dimensions}},\ }\href
  {https://doi.org/10.1103/PRXQuantum.6.010315} {\bibfield  {journal} {\bibinfo
   {journal} {PRX Quantum}\ }\textbf {\bibinfo {volume} {6}},\ \bibinfo {pages}
  {010315} (\bibinfo {year} {2025})}\BibitemShut {NoStop}%
\bibitem [{\citenamefont {Wang}\ \emph
  {et~al.}(2025{\natexlab{a}})\citenamefont {Wang}, \citenamefont {Wu},\ and\
  \citenamefont {Wang}}]{WangWuWang}%
  \BibitemOpen
  \bibfield  {author} {\bibinfo {author} {\bibfnamefont {Z.}~\bibnamefont
  {Wang}}, \bibinfo {author} {\bibfnamefont {Z.}~\bibnamefont {Wu}},\ and\
  \bibinfo {author} {\bibfnamefont {Z.}~\bibnamefont {Wang}},\ }\bibfield
  {title} {\bibinfo {title} {{Intrinsic Mixed-State Topological Order}},\
  }\href {https://doi.org/10.1103/PRXQuantum.6.010314} {\bibfield  {journal}
  {\bibinfo  {journal} {PRX Quantum}\ }\textbf {\bibinfo {volume} {6}},\
  \bibinfo {pages} {010314} (\bibinfo {year} {2025}{\natexlab{a}})}\BibitemShut
  {NoStop}%
\bibitem [{\citenamefont {Chen}\ and\ \citenamefont
  {Grover}(2024{\natexlab{a}})}]{YHChenGrover}%
  \BibitemOpen
  \bibfield  {author} {\bibinfo {author} {\bibfnamefont {Y.-H.}\ \bibnamefont
  {Chen}}\ and\ \bibinfo {author} {\bibfnamefont {T.}~\bibnamefont {Grover}},\
  }\bibfield  {title} {\bibinfo {title} {{Unconventional topological
  mixed-state transition and critical phase induced by self-dual coherent
  errors}},\ }\href {https://doi.org/10.1103/PhysRevB.110.125152} {\bibfield
  {journal} {\bibinfo  {journal} {Phys. Rev. B}\ }\textbf {\bibinfo {volume}
  {110}},\ \bibinfo {pages} {125152} (\bibinfo {year}
  {2024}{\natexlab{a}})}\BibitemShut {NoStop}%
\bibitem [{\citenamefont {Chen}\ and\ \citenamefont
  {Grover}(2024{\natexlab{b}})}]{YHChenGroverSeparability}%
  \BibitemOpen
  \bibfield  {author} {\bibinfo {author} {\bibfnamefont {Y.-H.}\ \bibnamefont
  {Chen}}\ and\ \bibinfo {author} {\bibfnamefont {T.}~\bibnamefont {Grover}},\
  }\bibfield  {title} {\bibinfo {title} {{Separability Transitions in
  Topological States Induced by Local Decoherence}},\ }\href
  {https://doi.org/10.1103/PhysRevLett.132.170602} {\bibfield  {journal}
  {\bibinfo  {journal} {Phys. Rev. Lett.}\ }\textbf {\bibinfo {volume} {132}},\
  \bibinfo {pages} {170602} (\bibinfo {year} {2024}{\natexlab{b}})}\BibitemShut
  {NoStop}%
\bibitem [{\citenamefont {Lessa}\ \emph {et~al.}(2025)\citenamefont {Lessa},
  \citenamefont {Ma}, \citenamefont {Zhang}, \citenamefont {Bi}, \citenamefont
  {Cheng},\ and\ \citenamefont {Wang}}]{LessaStrongToWeak}%
  \BibitemOpen
  \bibfield  {author} {\bibinfo {author} {\bibfnamefont {L.~A.}\ \bibnamefont
  {Lessa}}, \bibinfo {author} {\bibfnamefont {R.}~\bibnamefont {Ma}}, \bibinfo
  {author} {\bibfnamefont {J.-H.}\ \bibnamefont {Zhang}}, \bibinfo {author}
  {\bibfnamefont {Z.}~\bibnamefont {Bi}}, \bibinfo {author} {\bibfnamefont
  {M.}~\bibnamefont {Cheng}},\ and\ \bibinfo {author} {\bibfnamefont
  {C.}~\bibnamefont {Wang}},\ }\bibfield  {title} {\bibinfo {title}
  {{Strong-to-Weak Spontaneous Symmetry Breaking in Mixed Quantum States}},\
  }\href {https://doi.org/10.1103/PRXQuantum.6.010344} {\bibfield  {journal}
  {\bibinfo  {journal} {PRX Quantum}\ }\textbf {\bibinfo {volume} {6}},\
  \bibinfo {pages} {010344} (\bibinfo {year} {2025})}\BibitemShut {NoStop}%
\bibitem [{\citenamefont {Wang}\ \emph {et~al.}(2026)\citenamefont {Wang},
  \citenamefont {Kiely}, \citenamefont {Tell}, \citenamefont {Obermeyer},
  \citenamefont {Barendregt}, \citenamefont {Bojovi{\'c}}, \citenamefont
  {Preiss}, \citenamefont {Sarma}, \citenamefont {Franz}, \citenamefont
  {Fisher}, \citenamefont {Xu},\ and\ \citenamefont
  {Bloch}}]{WangKielyEtAl2026DephasedFermiGas}%
  \BibitemOpen
  \bibfield  {author} {\bibinfo {author} {\bibfnamefont {S.}~\bibnamefont
  {Wang}}, \bibinfo {author} {\bibfnamefont {T.~G.}\ \bibnamefont {Kiely}},
  \bibinfo {author} {\bibfnamefont {D.}~\bibnamefont {Tell}}, \bibinfo {author}
  {\bibfnamefont {J.}~\bibnamefont {Obermeyer}}, \bibinfo {author}
  {\bibfnamefont {M.}~\bibnamefont {Barendregt}}, \bibinfo {author}
  {\bibfnamefont {P.}~\bibnamefont {Bojovi{\'c}}}, \bibinfo {author}
  {\bibfnamefont {P.~M.}\ \bibnamefont {Preiss}}, \bibinfo {author}
  {\bibfnamefont {A.}~\bibnamefont {Sarma}}, \bibinfo {author} {\bibfnamefont
  {T.}~\bibnamefont {Franz}}, \bibinfo {author} {\bibfnamefont {M.~P.~A.}\
  \bibnamefont {Fisher}}, \bibinfo {author} {\bibfnamefont {C.}~\bibnamefont
  {Xu}},\ and\ \bibinfo {author} {\bibfnamefont {I.}~\bibnamefont {Bloch}},\
  }\href {https://doi.org/10.48550/arXiv.2604.16137} {\bibinfo {title}
  {{Observation of Strong-to-Weak Spontaneous Symmetry Breaking in a Dephased
  Fermi Gas}}} (\bibinfo {year} {2026}),\ \Eprint
  {https://arxiv.org/abs/2604.16137} {arXiv:2604.16137 [quant-ph]} \BibitemShut
  {NoStop}%
\bibitem [{\citenamefont {Sohal}\ and\ \citenamefont {Prem}(2025)}]{SohalPrem}%
  \BibitemOpen
  \bibfield  {author} {\bibinfo {author} {\bibfnamefont {R.}~\bibnamefont
  {Sohal}}\ and\ \bibinfo {author} {\bibfnamefont {A.}~\bibnamefont {Prem}},\
  }\bibfield  {title} {\bibinfo {title} {{Noisy Approach to Intrinsically
  Mixed-State Topological Order}},\ }\href
  {https://doi.org/10.1103/PRXQuantum.6.010313} {\bibfield  {journal} {\bibinfo
   {journal} {PRX Quantum}\ }\textbf {\bibinfo {volume} {6}},\ \bibinfo {pages}
  {010313} (\bibinfo {year} {2025})}\BibitemShut {NoStop}%
\bibitem [{\citenamefont {Su}\ \emph {et~al.}(2024)\citenamefont {Su},
  \citenamefont {Yang},\ and\ \citenamefont {Jian}}]{SuYangJian}%
  \BibitemOpen
  \bibfield  {author} {\bibinfo {author} {\bibfnamefont {K.}~\bibnamefont
  {Su}}, \bibinfo {author} {\bibfnamefont {Z.}~\bibnamefont {Yang}},\ and\
  \bibinfo {author} {\bibfnamefont {C.-M.}\ \bibnamefont {Jian}},\ }\bibfield
  {title} {\bibinfo {title} {{Tapestry of dualities in decohered quantum error
  correction codes}},\ }\href {https://doi.org/10.1103/PhysRevB.110.085158}
  {\bibfield  {journal} {\bibinfo  {journal} {Phys. Rev. B}\ }\textbf {\bibinfo
  {volume} {110}},\ \bibinfo {pages} {085158} (\bibinfo {year}
  {2024})}\BibitemShut {NoStop}%
\bibitem [{\citenamefont {Lee}(2025)}]{LeeExactCalculation}%
  \BibitemOpen
  \bibfield  {author} {\bibinfo {author} {\bibfnamefont {J.~Y.}\ \bibnamefont
  {Lee}},\ }\bibfield  {title} {\bibinfo {title} {{Exact Calculations of
  Coherent Information for Toric Codes under Decoherence: Identifying the
  Fundamental Error Threshold}},\ }\href {https://doi.org/10.1103/hlfh-86yz}
  {\bibfield  {journal} {\bibinfo  {journal} {Phys. Rev. Lett.}\ }\textbf
  {\bibinfo {volume} {134}},\ \bibinfo {pages} {250601} (\bibinfo {year}
  {2025})}\BibitemShut {NoStop}%
\bibitem [{\citenamefont {Eckstein}\ \emph {et~al.}(2024)\citenamefont
  {Eckstein}, \citenamefont {Han}, \citenamefont {Trebst},\ and\ \citenamefont
  {Zhu}}]{EcksteinPRX}%
  \BibitemOpen
  \bibfield  {author} {\bibinfo {author} {\bibfnamefont {F.}~\bibnamefont
  {Eckstein}}, \bibinfo {author} {\bibfnamefont {B.}~\bibnamefont {Han}},
  \bibinfo {author} {\bibfnamefont {S.}~\bibnamefont {Trebst}},\ and\ \bibinfo
  {author} {\bibfnamefont {G.-Y.}\ \bibnamefont {Zhu}},\ }\bibfield  {title}
  {\bibinfo {title} {{Robust Teleportation of a Surface Code and Cascade of
  Topological Quantum Phase Transitions}},\ }\href
  {https://doi.org/10.1103/PRXQuantum.5.040313} {\bibfield  {journal} {\bibinfo
   {journal} {PRX Quantum}\ }\textbf {\bibinfo {volume} {5}},\ \bibinfo {pages}
  {040313} (\bibinfo {year} {2024})}\BibitemShut {NoStop}%
\bibitem [{\citenamefont {Wang}\ \emph
  {et~al.}(2025{\natexlab{b}})\citenamefont {Wang}, \citenamefont {Vasseur},
  \citenamefont {Trebst}, \citenamefont {Ludwig},\ and\ \citenamefont
  {Zhu}}]{DecoherenceWangEtAl}%
  \BibitemOpen
  \bibfield  {author} {\bibinfo {author} {\bibfnamefont {Q.}~\bibnamefont
  {Wang}}, \bibinfo {author} {\bibfnamefont {R.}~\bibnamefont {Vasseur}},
  \bibinfo {author} {\bibfnamefont {S.}~\bibnamefont {Trebst}}, \bibinfo
  {author} {\bibfnamefont {A.~W.~W.}\ \bibnamefont {Ludwig}},\ and\ \bibinfo
  {author} {\bibfnamefont {G.-Y.}\ \bibnamefont {Zhu}},\ }\href@noop {}
  {\bibinfo {title} {{Decoherence-induced self-dual criticality in topological
  states of matter}}} (\bibinfo {year} {2025}{\natexlab{b}}),\ \Eprint
  {https://arxiv.org/abs/2502.14034} {arXiv:2502.14034 [quant-ph]} \BibitemShut
  {NoStop}%
\bibitem [{\citenamefont {Eckstein}\ \emph {et~al.}(2025)\citenamefont
  {Eckstein}, \citenamefont {Han}, \citenamefont {Trebst},\ and\ \citenamefont
  {Zhu}}]{eckstein2025learningtransitionstopologicalsurface}%
  \BibitemOpen
  \bibfield  {author} {\bibinfo {author} {\bibfnamefont {F.}~\bibnamefont
  {Eckstein}}, \bibinfo {author} {\bibfnamefont {B.}~\bibnamefont {Han}},
  \bibinfo {author} {\bibfnamefont {S.}~\bibnamefont {Trebst}},\ and\ \bibinfo
  {author} {\bibfnamefont {G.-Y.}\ \bibnamefont {Zhu}},\ }\href
  {https://arxiv.org/abs/2512.19786} {\bibinfo {title} {{Learning transitions
  of topological surface codes}}} (\bibinfo {year} {2025}),\ \Eprint
  {https://arxiv.org/abs/2512.19786} {arXiv:2512.19786 [quant-ph]} \BibitemShut
  {NoStop}%
\bibitem [{\citenamefont {Huang}\ \emph {et~al.}(2025)\citenamefont {Huang},
  \citenamefont {Colmenarez}, \citenamefont {M\"uller},\ and\ \citenamefont
  {Diehl}}]{DiehlEtAl2025}%
  \BibitemOpen
  \bibfield  {author} {\bibinfo {author} {\bibfnamefont {Z.-M.}\ \bibnamefont
  {Huang}}, \bibinfo {author} {\bibfnamefont {L.}~\bibnamefont {Colmenarez}},
  \bibinfo {author} {\bibfnamefont {M.}~\bibnamefont {M\"uller}},\ and\
  \bibinfo {author} {\bibfnamefont {S.}~\bibnamefont {Diehl}},\ }\bibfield
  {title} {\bibinfo {title} {{Coherent information as a mixed-state topological
  order parameter of fermions}},\ }\href {https://doi.org/10.1103/fx56-8nvy}
  {\bibfield  {journal} {\bibinfo  {journal} {Phys. Rev. Res.}\ }\textbf
  {\bibinfo {volume} {7}},\ \bibinfo {pages} {043009} (\bibinfo {year}
  {2025})}\BibitemShut {NoStop}%
\bibitem [{\citenamefont {Wan}\ \emph {et~al.}(2025)\citenamefont {Wan},
  \citenamefont {Dai},\ and\ \citenamefont
  {Zhu}}]{wan2025revisitingnishimorimulticriticalitylens}%
  \BibitemOpen
  \bibfield  {author} {\bibinfo {author} {\bibfnamefont {Z.-Q.}\ \bibnamefont
  {Wan}}, \bibinfo {author} {\bibfnamefont {X.-D.}\ \bibnamefont {Dai}},\ and\
  \bibinfo {author} {\bibfnamefont {G.-Y.}\ \bibnamefont {Zhu}},\ }\href
  {https://arxiv.org/abs/2511.02907} {\bibinfo {title} {{Revisiting Nishimori
  multicriticality through the lens of information measures}}} (\bibinfo {year}
  {2025}),\ \Eprint {https://arxiv.org/abs/2511.02907} {arXiv:2511.02907
  [cond-mat.stat-mech]} \BibitemShut {NoStop}%
\bibitem [{\citenamefont {Garratt}\ \emph {et~al.}(2023)\citenamefont
  {Garratt}, \citenamefont {Weinstein},\ and\ \citenamefont
  {Altman}}]{Garratt22}%
  \BibitemOpen
  \bibfield  {author} {\bibinfo {author} {\bibfnamefont {S.~J.}\ \bibnamefont
  {Garratt}}, \bibinfo {author} {\bibfnamefont {Z.}~\bibnamefont {Weinstein}},\
  and\ \bibinfo {author} {\bibfnamefont {E.}~\bibnamefont {Altman}},\
  }\bibfield  {title} {\bibinfo {title} {{Measurements Conspire Nonlocally to
  Restructure Critical Quantum States}},\ }\href
  {https://doi.org/10.1103/PhysRevX.13.021026} {\bibfield  {journal} {\bibinfo
  {journal} {Phys. Rev. X}\ }\textbf {\bibinfo {volume} {13}},\ \bibinfo
  {pages} {021026} (\bibinfo {year} {2023})}\BibitemShut {NoStop}%
\bibitem [{\citenamefont {Weinstein}\ \emph {et~al.}(2023)\citenamefont
  {Weinstein}, \citenamefont {Sajith}, \citenamefont {Altman},\ and\
  \citenamefont {Garratt}}]{Garratt23measureising}%
  \BibitemOpen
  \bibfield  {author} {\bibinfo {author} {\bibfnamefont {Z.}~\bibnamefont
  {Weinstein}}, \bibinfo {author} {\bibfnamefont {R.}~\bibnamefont {Sajith}},
  \bibinfo {author} {\bibfnamefont {E.}~\bibnamefont {Altman}},\ and\ \bibinfo
  {author} {\bibfnamefont {S.~J.}\ \bibnamefont {Garratt}},\ }\bibfield
  {title} {\bibinfo {title} {{Nonlocality and entanglement in measured critical
  quantum Ising chains}},\ }\href {https://doi.org/10.1103/PhysRevB.107.245132}
  {\bibfield  {journal} {\bibinfo  {journal} {Phys. Rev. B}\ }\textbf {\bibinfo
  {volume} {107}},\ \bibinfo {pages} {245132} (\bibinfo {year}
  {2023})}\BibitemShut {NoStop}%
\bibitem [{\citenamefont {Murciano}\ \emph {et~al.}(2023)\citenamefont
  {Murciano}, \citenamefont {Sala}, \citenamefont {Liu}, \citenamefont {Mong},\
  and\ \citenamefont {Alicea}}]{Alicea23measureising}%
  \BibitemOpen
  \bibfield  {author} {\bibinfo {author} {\bibfnamefont {S.}~\bibnamefont
  {Murciano}}, \bibinfo {author} {\bibfnamefont {P.}~\bibnamefont {Sala}},
  \bibinfo {author} {\bibfnamefont {Y.}~\bibnamefont {Liu}}, \bibinfo {author}
  {\bibfnamefont {R.~S.~K.}\ \bibnamefont {Mong}},\ and\ \bibinfo {author}
  {\bibfnamefont {J.}~\bibnamefont {Alicea}},\ }\bibfield  {title} {\bibinfo
  {title} {{Measurement-Altered Ising Quantum Criticality}},\ }\href
  {https://doi.org/10.1103/PhysRevX.13.041042} {\bibfield  {journal} {\bibinfo
  {journal} {Phys. Rev. X}\ }\textbf {\bibinfo {volume} {13}},\ \bibinfo
  {pages} {041042} (\bibinfo {year} {2023})}\BibitemShut {NoStop}%
\bibitem [{\citenamefont {Sala}\ \emph
  {et~al.}(2024{\natexlab{b}})\citenamefont {Sala}, \citenamefont {Murciano},
  \citenamefont {Liu},\ and\ \citenamefont {Alicea}}]{Alicea24teleportation}%
  \BibitemOpen
  \bibfield  {author} {\bibinfo {author} {\bibfnamefont {P.}~\bibnamefont
  {Sala}}, \bibinfo {author} {\bibfnamefont {S.}~\bibnamefont {Murciano}},
  \bibinfo {author} {\bibfnamefont {Y.}~\bibnamefont {Liu}},\ and\ \bibinfo
  {author} {\bibfnamefont {J.}~\bibnamefont {Alicea}},\ }\bibfield  {title}
  {\bibinfo {title} {{Quantum Criticality Under Imperfect Teleportation}},\
  }\href {https://doi.org/10.1103/PRXQuantum.5.030307} {\bibfield  {journal}
  {\bibinfo  {journal} {PRX Quantum}\ }\textbf {\bibinfo {volume} {5}},\
  \bibinfo {pages} {030307} (\bibinfo {year} {2024}{\natexlab{b}})}\BibitemShut
  {NoStop}%
\bibitem [{\citenamefont {Yang}\ \emph {et~al.}(2023)\citenamefont {Yang},
  \citenamefont {Mao},\ and\ \citenamefont {Jian}}]{jian23measureising}%
  \BibitemOpen
  \bibfield  {author} {\bibinfo {author} {\bibfnamefont {Z.}~\bibnamefont
  {Yang}}, \bibinfo {author} {\bibfnamefont {D.}~\bibnamefont {Mao}},\ and\
  \bibinfo {author} {\bibfnamefont {C.-M.}\ \bibnamefont {Jian}},\ }\bibfield
  {title} {\bibinfo {title} {{Entanglement in a one-dimensional critical state
  after measurements}},\ }\href {https://doi.org/10.1103/PhysRevB.108.165120}
  {\bibfield  {journal} {\bibinfo  {journal} {Phys. Rev. B}\ }\textbf {\bibinfo
  {volume} {108}},\ \bibinfo {pages} {165120} (\bibinfo {year}
  {2023})}\BibitemShut {NoStop}%
\bibitem [{\citenamefont {{Patil}}\ and\ \citenamefont
  {{Ludwig}}(2024)}]{Ludwig24measurecritical}%
  \BibitemOpen
  \bibfield  {author} {\bibinfo {author} {\bibfnamefont {R.~A.}\ \bibnamefont
  {{Patil}}}\ and\ \bibinfo {author} {\bibfnamefont {A.~W.~W.}\ \bibnamefont
  {{Ludwig}}},\ }\bibfield  {title} {\bibinfo {title} {{Highly complex novel
  critical behavior from the intrinsic randomness of quantum mechanical
  measurements on critical ground states -- a controlled renormalization group
  analysis}},\ }\href@noop {} {\bibfield  {journal} {\bibinfo  {journal}
  {preprint}\ } (\bibinfo {year} {2024})},\ \Eprint
  {https://arxiv.org/abs/2409.02107} {arXiv:2409.02107} \BibitemShut {NoStop}%
\bibitem [{\citenamefont {Liu}\ \emph {et~al.}(2025{\natexlab{a}})\citenamefont
  {Liu}, \citenamefont {Murciano}, \citenamefont {Mross},\ and\ \citenamefont
  {Alicea}}]{liu2024boundary}%
  \BibitemOpen
  \bibfield  {author} {\bibinfo {author} {\bibfnamefont {Y.}~\bibnamefont
  {Liu}}, \bibinfo {author} {\bibfnamefont {S.}~\bibnamefont {Murciano}},
  \bibinfo {author} {\bibfnamefont {D.~F.}\ \bibnamefont {Mross}},\ and\
  \bibinfo {author} {\bibfnamefont {J.}~\bibnamefont {Alicea}},\ }\bibfield
  {title} {\bibinfo {title} {{Boundary transitions from a single round of
  measurements on gapless quantum states}},\ }\href
  {https://doi.org/10.1103/PhysRevResearch.7.023293} {\bibfield  {journal}
  {\bibinfo  {journal} {Phys. Rev. Res.}\ }\textbf {\bibinfo {volume} {7}},\
  \bibinfo {pages} {023293} (\bibinfo {year} {2025}{\natexlab{a}})}\BibitemShut
  {NoStop}%
\bibitem [{\citenamefont {Baweja}\ \emph {et~al.}(2025)\citenamefont {Baweja},
  \citenamefont {Luitz},\ and\ \citenamefont
  {Garratt}}]{LuitzGarratt2024MeasStatesHigherDim}%
  \BibitemOpen
  \bibfield  {author} {\bibinfo {author} {\bibfnamefont {K.}~\bibnamefont
  {Baweja}}, \bibinfo {author} {\bibfnamefont {D.~J.}\ \bibnamefont {Luitz}},\
  and\ \bibinfo {author} {\bibfnamefont {S.~J.}\ \bibnamefont {Garratt}},\
  }\bibfield  {title} {\bibinfo {title} {{Post-measurement Quantum Monte
  Carlo}},\ }\href {https://doi.org/10.1103/PhysRevB.112.184417} {\bibfield
  {journal} {\bibinfo  {journal} {Phys. Rev. B}\ }\textbf {\bibinfo {volume}
  {112}},\ \bibinfo {pages} {184417} (\bibinfo {year} {2025})}\BibitemShut
  {NoStop}%
\bibitem [{\citenamefont {Hoshino}\ \emph {et~al.}(2025)\citenamefont
  {Hoshino}, \citenamefont {Oshikawa},\ and\ \citenamefont
  {Ashida}}]{hoshino2024entanglement}%
  \BibitemOpen
  \bibfield  {author} {\bibinfo {author} {\bibfnamefont {M.}~\bibnamefont
  {Hoshino}}, \bibinfo {author} {\bibfnamefont {M.}~\bibnamefont {Oshikawa}},\
  and\ \bibinfo {author} {\bibfnamefont {Y.}~\bibnamefont {Ashida}},\
  }\bibfield  {title} {\bibinfo {title} {Entanglement swapping in critical
  quantum spin chains},\ }\href {https://doi.org/10.1103/PhysRevB.111.155143}
  {\bibfield  {journal} {\bibinfo  {journal} {Phys. Rev. B}\ }\textbf {\bibinfo
  {volume} {111}},\ \bibinfo {pages} {155143} (\bibinfo {year}
  {2025})}\BibitemShut {NoStop}%
\bibitem [{\citenamefont {Tang}\ and\ \citenamefont
  {Wen}(2024)}]{tang2024critical}%
  \BibitemOpen
  \bibfield  {author} {\bibinfo {author} {\bibfnamefont {Q.}~\bibnamefont
  {Tang}}\ and\ \bibinfo {author} {\bibfnamefont {X.}~\bibnamefont {Wen}},\
  }\bibfield  {title} {\bibinfo {title} {A critical state under weak
  measurement is not critical},\ }\href@noop {} {\bibfield  {journal} {\bibinfo
   {journal} {preprint}\ } (\bibinfo {year} {2024})},\ \Eprint
  {https://arxiv.org/abs/2411.13705} {arXiv:2411.13705} \BibitemShut {NoStop}%
\bibitem [{\citenamefont {P{\"u}tz}\ \emph {et~al.}(2025)\citenamefont
  {P{\"u}tz}, \citenamefont {Garratt}, \citenamefont {Nishimori}, \citenamefont
  {Trebst},\ and\ \citenamefont {Zhu}}]{PuetzGarrattNishimoriTrebstZhu2025}%
  \BibitemOpen
  \bibfield  {author} {\bibinfo {author} {\bibfnamefont {M.}~\bibnamefont
  {P{\"u}tz}}, \bibinfo {author} {\bibfnamefont {S.~J.}\ \bibnamefont
  {Garratt}}, \bibinfo {author} {\bibfnamefont {H.}~\bibnamefont {Nishimori}},
  \bibinfo {author} {\bibfnamefont {S.}~\bibnamefont {Trebst}},\ and\ \bibinfo
  {author} {\bibfnamefont {G.-Y.}\ \bibnamefont {Zhu}},\ }\href
  {https://doi.org/10.48550/arXiv.2504.12385} {\bibinfo {title} {{Learning
  transitions in classical Ising models and deformed toric codes}}} (\bibinfo
  {year} {2025}),\ \Eprint {https://arxiv.org/abs/2504.12385} {arXiv:2504.12385
  [cond-mat.stat-mech]} \BibitemShut {NoStop}%
\bibitem [{\citenamefont {Patil}\ \emph {et~al.}(2026)\citenamefont {Patil},
  \citenamefont {P{\"u}tz}, \citenamefont {Trebst}, \citenamefont {Zhu},\ and\
  \citenamefont {Ludwig}}]{PatilPuetzTrebstZhuLudwig2026}%
  \BibitemOpen
  \bibfield  {author} {\bibinfo {author} {\bibfnamefont {R.~A.}\ \bibnamefont
  {Patil}}, \bibinfo {author} {\bibfnamefont {M.}~\bibnamefont {P{\"u}tz}},
  \bibinfo {author} {\bibfnamefont {S.}~\bibnamefont {Trebst}}, \bibinfo
  {author} {\bibfnamefont {G.-Y.}\ \bibnamefont {Zhu}},\ and\ \bibinfo {author}
  {\bibfnamefont {A.~W.~W.}\ \bibnamefont {Ludwig}},\ }\href
  {https://doi.org/10.48550/arXiv.2604.06324} {\bibinfo {title} {{Higher
  Nishimori Criticality and Exact Results at the Learning Transition of
  Deformed Toric Codes}}} (\bibinfo {year} {2026}),\ \Eprint
  {https://arxiv.org/abs/2604.06324} {arXiv:2604.06324 [cond-mat.stat-mech]}
  \BibitemShut {NoStop}%
\bibitem [{\citenamefont {Polkovnikov}(2011)}]{POLKOVNIKOV2011486}%
  \BibitemOpen
  \bibfield  {author} {\bibinfo {author} {\bibfnamefont {A.}~\bibnamefont
  {Polkovnikov}},\ }\bibfield  {title} {\bibinfo {title} {Microscopic diagonal
  entropy and its connection to basic thermodynamic relations},\ }\href
  {https://doi.org/https://doi.org/10.1016/j.aop.2010.08.004} {\bibfield
  {journal} {\bibinfo  {journal} {Annals of Physics}\ }\textbf {\bibinfo
  {volume} {326}},\ \bibinfo {pages} {486} (\bibinfo {year}
  {2011})}\BibitemShut {NoStop}%
\bibitem [{\citenamefont {Chen}\ and\ \citenamefont
  {Grover}(2025)}]{ChenGrover2025Zipping}%
  \BibitemOpen
  \bibfield  {author} {\bibinfo {author} {\bibfnamefont {Y.-H.}\ \bibnamefont
  {Chen}}\ and\ \bibinfo {author} {\bibfnamefont {T.}~\bibnamefont {Grover}},\
  }\href {https://arxiv.org/abs/2502.18898} {\bibinfo {title} {{Zipping
  many-body quantum states: a scalable approach to diagonal entropy}}}
  (\bibinfo {year} {2025}),\ \Eprint {https://arxiv.org/abs/2502.18898}
  {arXiv:2502.18898 [quant-ph]} \BibitemShut {NoStop}%
\bibitem [{\citenamefont {Heyl}\ and\ \citenamefont
  {Budich}(2017)}]{PhysRevB.96.180304}%
  \BibitemOpen
  \bibfield  {author} {\bibinfo {author} {\bibfnamefont {M.}~\bibnamefont
  {Heyl}}\ and\ \bibinfo {author} {\bibfnamefont {J.~C.}\ \bibnamefont
  {Budich}},\ }\bibfield  {title} {\bibinfo {title} {Dynamical topological
  quantum phase transitions for mixed states},\ }\href
  {https://doi.org/10.1103/PhysRevB.96.180304} {\bibfield  {journal} {\bibinfo
  {journal} {Phys. Rev. B}\ }\textbf {\bibinfo {volume} {96}},\ \bibinfo
  {pages} {180304} (\bibinfo {year} {2017})}\BibitemShut {NoStop}%
\bibitem [{\citenamefont {Bhattacharya}\ \emph {et~al.}(2017)\citenamefont
  {Bhattacharya}, \citenamefont {Bandyopadhyay},\ and\ \citenamefont
  {Dutta}}]{PhysRevB.96.180303}%
  \BibitemOpen
  \bibfield  {author} {\bibinfo {author} {\bibfnamefont {U.}~\bibnamefont
  {Bhattacharya}}, \bibinfo {author} {\bibfnamefont {S.}~\bibnamefont
  {Bandyopadhyay}},\ and\ \bibinfo {author} {\bibfnamefont {A.}~\bibnamefont
  {Dutta}},\ }\bibfield  {title} {\bibinfo {title} {Mixed state dynamical
  quantum phase transitions},\ }\href
  {https://doi.org/10.1103/PhysRevB.96.180303} {\bibfield  {journal} {\bibinfo
  {journal} {Phys. Rev. B}\ }\textbf {\bibinfo {volume} {96}},\ \bibinfo
  {pages} {180303} (\bibinfo {year} {2017})}\BibitemShut {NoStop}%
\bibitem [{\citenamefont {Lang}\ \emph {et~al.}(2018)\citenamefont {Lang},
  \citenamefont {Chen}, \citenamefont {Hong},\ and\ \citenamefont
  {Fan}}]{PhysRevB.98.134310}%
  \BibitemOpen
  \bibfield  {author} {\bibinfo {author} {\bibfnamefont {H.}~\bibnamefont
  {Lang}}, \bibinfo {author} {\bibfnamefont {Y.}~\bibnamefont {Chen}}, \bibinfo
  {author} {\bibfnamefont {Q.}~\bibnamefont {Hong}},\ and\ \bibinfo {author}
  {\bibfnamefont {H.}~\bibnamefont {Fan}},\ }\bibfield  {title} {\bibinfo
  {title} {Dynamical quantum phase transition for mixed states in open
  systems},\ }\href {https://doi.org/10.1103/PhysRevB.98.134310} {\bibfield
  {journal} {\bibinfo  {journal} {Phys. Rev. B}\ }\textbf {\bibinfo {volume}
  {98}},\ \bibinfo {pages} {134310} (\bibinfo {year} {2018})}\BibitemShut
  {NoStop}%
\bibitem [{\citenamefont {Fishman}\ \emph {et~al.}(2022)\citenamefont
  {Fishman}, \citenamefont {White},\ and\ \citenamefont
  {Stoudenmire}}]{ITensor}%
  \BibitemOpen
  \bibfield  {author} {\bibinfo {author} {\bibfnamefont {M.}~\bibnamefont
  {Fishman}}, \bibinfo {author} {\bibfnamefont {S.~R.}\ \bibnamefont {White}},\
  and\ \bibinfo {author} {\bibfnamefont {E.~M.}\ \bibnamefont {Stoudenmire}},\
  }\bibfield  {title} {\bibinfo {title} {{The ITensor Software Library for
  Tensor Network Calculations}},\ }\href
  {https://doi.org/10.21468/SciPostPhysCodeb.4} {\bibfield  {journal} {\bibinfo
   {journal} {SciPost Phys. Codebases}\ ,\ \bibinfo {pages} {4}} (\bibinfo
  {year} {2022})}\BibitemShut {NoStop}%
\bibitem [{\citenamefont {Guo}\ and\ \citenamefont {Zhu}()}]{GuoZhu26fermion}%
  \BibitemOpen
  \bibfield  {author} {\bibinfo {author} {\bibfnamefont {R.-J.}\ \bibnamefont
  {Guo}}\ and\ \bibinfo {author} {\bibfnamefont {G.-Y.}\ \bibnamefont {Zhu}},\
  }\href@noop {} {\bibinfo {title} {{To appear}}}\BibitemShut {NoStop}%
\bibitem [{Note1()}]{Note1}%
  \BibitemOpen
  \bibinfo {note} {A comment on the numerical method: the probability can be
  viewed as the amplitude of a time evolved Gaussian fermion state in the Fock
  space, by Jordan Wigner transforming the Ising Hamiltonian to the free
  fermion Hamiltonian in the parity even sector $\DOTSB \tprod \slimits@ _j
  X_j=+1$, which maps the periodic boundary condition of the spin chain to the
  anti-periodic boundary condition of the fermion chain. Using fermion basis
  saves the need to numerically diagonalizing the Hamiltonian or trotterizing
  the Hamiltonian. However, the numerical computation is limited by enumerating
  the measurement outcome configurations, especially if we want to analyze the
  multifractal structure with small moment $q$ that emphasizes the less
  probable configurations. Since in fermion approach one stores an $2L$-by-$2L$
  matrix that captures the Gaussian fermion state, one has to enumerate $2^L$
  times of computation to calculate the whole probability distribution
  function. Similarly for tensor network approximation approach, where the
  spatial complexity of numerical computation is polynomial but we still
  require exponential time complexity to extract the whole distribution
  function. Therefore for most calculation we adopt exact state evolution by a
  trotterized Hamiltonian, with exponential overhead in the spatial complexity
  but not in time complexity. We also leave a discussion of numerical sampling
  in the appendix.}\BibitemShut {Stop}%
\bibitem [{\citenamefont {Wong}\ and\ \citenamefont
  {Yu}(2022)}]{PhysRevB.105.174307}%
  \BibitemOpen
  \bibfield  {author} {\bibinfo {author} {\bibfnamefont {C.~Y.}\ \bibnamefont
  {Wong}}\ and\ \bibinfo {author} {\bibfnamefont {W.~C.}\ \bibnamefont {Yu}},\
  }\bibfield  {title} {\bibinfo {title} {Loschmidt amplitude spectrum in
  dynamical quantum phase transitions},\ }\href
  {https://doi.org/10.1103/PhysRevB.105.174307} {\bibfield  {journal} {\bibinfo
   {journal} {Phys. Rev. B}\ }\textbf {\bibinfo {volume} {105}},\ \bibinfo
  {pages} {174307} (\bibinfo {year} {2022})}\BibitemShut {NoStop}%
\bibitem [{\citenamefont {Zabalo}\ \emph {et~al.}(2022)\citenamefont {Zabalo},
  \citenamefont {Gullans}, \citenamefont {Wilson}, \citenamefont {Vasseur},
  \citenamefont {Ludwig}, \citenamefont {Gopalakrishnan}, \citenamefont
  {Huse},\ and\ \citenamefont
  {Pixley}}]{ZabaloGullansWilsonVasseurLudwigGopalakrishnanHusePixley}%
  \BibitemOpen
  \bibfield  {author} {\bibinfo {author} {\bibfnamefont {A.}~\bibnamefont
  {Zabalo}}, \bibinfo {author} {\bibfnamefont {M.~J.}\ \bibnamefont {Gullans}},
  \bibinfo {author} {\bibfnamefont {J.~H.}\ \bibnamefont {Wilson}}, \bibinfo
  {author} {\bibfnamefont {R.}~\bibnamefont {Vasseur}}, \bibinfo {author}
  {\bibfnamefont {A.~W.~W.}\ \bibnamefont {Ludwig}}, \bibinfo {author}
  {\bibfnamefont {S.}~\bibnamefont {Gopalakrishnan}}, \bibinfo {author}
  {\bibfnamefont {D.~A.}\ \bibnamefont {Huse}},\ and\ \bibinfo {author}
  {\bibfnamefont {J.~H.}\ \bibnamefont {Pixley}},\ }\bibfield  {title}
  {\bibinfo {title} {{Operator Scaling Dimensions and Multifractality at
  Measurement-Induced Transitions}},\ }\href
  {https://doi.org/10.1103/PhysRevLett.128.050602} {\bibfield  {journal}
  {\bibinfo  {journal} {Phys. Rev. Lett.}\ }\textbf {\bibinfo {volume} {128}},\
  \bibinfo {pages} {050602} (\bibinfo {year} {2022})}\BibitemShut {NoStop}%
\bibitem [{\citenamefont {P\"utz}\ \emph {et~al.}(2025)\citenamefont {P\"utz},
  \citenamefont {Vasseur}, \citenamefont {Ludwig}, \citenamefont {Trebst},\
  and\ \citenamefont {Zhu}}]{putz2025flownishimoriuniversalityweakly}%
  \BibitemOpen
  \bibfield  {author} {\bibinfo {author} {\bibfnamefont {M.}~\bibnamefont
  {P\"utz}}, \bibinfo {author} {\bibfnamefont {R.}~\bibnamefont {Vasseur}},
  \bibinfo {author} {\bibfnamefont {A.~W.}\ \bibnamefont {Ludwig}}, \bibinfo
  {author} {\bibfnamefont {S.}~\bibnamefont {Trebst}},\ and\ \bibinfo {author}
  {\bibfnamefont {G.-Y.}\ \bibnamefont {Zhu}},\ }\bibfield  {title} {\bibinfo
  {title} {{Flow to Nishimori Universality in Weakly Monitored Quantum Circuits
  with Qubit Loss}},\ }\href {https://doi.org/10.1103/ygfz-crvp} {\bibfield
  {journal} {\bibinfo  {journal} {PRX Quantum}\ }\textbf {\bibinfo {volume}
  {6}},\ \bibinfo {pages} {040372} (\bibinfo {year} {2025})}\BibitemShut
  {NoStop}%
\bibitem [{\citenamefont {Evers}\ and\ \citenamefont
  {Mirlin}(2008)}]{Mirlin08rmp}%
  \BibitemOpen
  \bibfield  {author} {\bibinfo {author} {\bibfnamefont {F.}~\bibnamefont
  {Evers}}\ and\ \bibinfo {author} {\bibfnamefont {A.~D.}\ \bibnamefont
  {Mirlin}},\ }\bibfield  {title} {\bibinfo {title} {Anderson transitions},\
  }\href {https://doi.org/10.1103/RevModPhys.80.1355} {\bibfield  {journal}
  {\bibinfo  {journal} {Rev. Mod. Phys.}\ }\textbf {\bibinfo {volume} {80}},\
  \bibinfo {pages} {1355} (\bibinfo {year} {2008})}\BibitemShut {NoStop}%
\bibitem [{\citenamefont {Mac\'e}\ \emph {et~al.}(2019)\citenamefont {Mac\'e},
  \citenamefont {Alet},\ and\ \citenamefont
  {Laflorencie}}]{PhysRevLett.123.180601}%
  \BibitemOpen
  \bibfield  {author} {\bibinfo {author} {\bibfnamefont {N.}~\bibnamefont
  {Mac\'e}}, \bibinfo {author} {\bibfnamefont {F.}~\bibnamefont {Alet}},\ and\
  \bibinfo {author} {\bibfnamefont {N.}~\bibnamefont {Laflorencie}},\
  }\bibfield  {title} {\bibinfo {title} {Multifractal scalings across the
  many-body localization transition},\ }\href
  {https://doi.org/10.1103/PhysRevLett.123.180601} {\bibfield  {journal}
  {\bibinfo  {journal} {Phys. Rev. Lett.}\ }\textbf {\bibinfo {volume} {123}},\
  \bibinfo {pages} {180601} (\bibinfo {year} {2019})}\BibitemShut {NoStop}%
\bibitem [{\citenamefont {Liu}\ \emph {et~al.}(2025{\natexlab{b}})\citenamefont
  {Liu}, \citenamefont {Sierant}, \citenamefont {Stornati}, \citenamefont
  {Lewenstein},\ and\ \citenamefont {P\l{}odzie\ifmmode~\acute{n}\else
  \'{n}\fi{}}}]{PhysRevA.111.052614}%
  \BibitemOpen
  \bibfield  {author} {\bibinfo {author} {\bibfnamefont {Y.}~\bibnamefont
  {Liu}}, \bibinfo {author} {\bibfnamefont {P.}~\bibnamefont {Sierant}},
  \bibinfo {author} {\bibfnamefont {P.}~\bibnamefont {Stornati}}, \bibinfo
  {author} {\bibfnamefont {M.}~\bibnamefont {Lewenstein}},\ and\ \bibinfo
  {author} {\bibfnamefont {M.}~\bibnamefont {P\l{}odzie\ifmmode~\acute{n}\else
  \'{n}\fi{}}},\ }\bibfield  {title} {\bibinfo {title} {Quantum algorithms for
  inverse participation ratio estimation in multiqubit and multiqudit
  systems},\ }\href {https://doi.org/10.1103/PhysRevA.111.052614} {\bibfield
  {journal} {\bibinfo  {journal} {Phys. Rev. A}\ }\textbf {\bibinfo {volume}
  {111}},\ \bibinfo {pages} {052614} (\bibinfo {year}
  {2025}{\natexlab{b}})}\BibitemShut {NoStop}%
\bibitem [{\citenamefont {Jian}\ \emph {et~al.}(2020)\citenamefont {Jian},
  \citenamefont {You}, \citenamefont {Vasseur},\ and\ \citenamefont
  {Ludwig}}]{Ludwig2020}%
  \BibitemOpen
  \bibfield  {author} {\bibinfo {author} {\bibfnamefont {C.-M.}\ \bibnamefont
  {Jian}}, \bibinfo {author} {\bibfnamefont {Y.-Z.}\ \bibnamefont {You}},
  \bibinfo {author} {\bibfnamefont {R.}~\bibnamefont {Vasseur}},\ and\ \bibinfo
  {author} {\bibfnamefont {A.~W.~W.}\ \bibnamefont {Ludwig}},\ }\bibfield
  {title} {\bibinfo {title} {{Measurement-induced criticality in random quantum
  circuits}},\ }\href {https://doi.org/10.1103/PhysRevB.101.104302} {\bibfield
  {journal} {\bibinfo  {journal} {Phys. Rev. B}\ }\textbf {\bibinfo {volume}
  {101}},\ \bibinfo {pages} {104302} (\bibinfo {year} {2020})}\BibitemShut
  {NoStop}%
\bibitem [{\citenamefont {Silva}(2008)}]{PhysRevLett.101.120603}%
  \BibitemOpen
  \bibfield  {author} {\bibinfo {author} {\bibfnamefont {A.}~\bibnamefont
  {Silva}},\ }\bibfield  {title} {\bibinfo {title} {Statistics of the work done
  on a quantum critical system by quenching a control parameter},\ }\href
  {https://doi.org/10.1103/PhysRevLett.101.120603} {\bibfield  {journal}
  {\bibinfo  {journal} {Phys. Rev. Lett.}\ }\textbf {\bibinfo {volume} {101}},\
  \bibinfo {pages} {120603} (\bibinfo {year} {2008})}\BibitemShut {NoStop}%
\bibitem [{\citenamefont {Briegel}\ \emph {et~al.}(2009)\citenamefont
  {Briegel}, \citenamefont {Browne}, \citenamefont {D{\"u}r}, \citenamefont
  {Raussendorf},\ and\ \citenamefont {Van~den Nest}}]{Briegel2009}%
  \BibitemOpen
  \bibfield  {author} {\bibinfo {author} {\bibfnamefont {H.~J.}\ \bibnamefont
  {Briegel}}, \bibinfo {author} {\bibfnamefont {D.~E.}\ \bibnamefont {Browne}},
  \bibinfo {author} {\bibfnamefont {W.}~\bibnamefont {D{\"u}r}}, \bibinfo
  {author} {\bibfnamefont {R.}~\bibnamefont {Raussendorf}},\ and\ \bibinfo
  {author} {\bibfnamefont {M.}~\bibnamefont {Van~den Nest}},\ }\bibfield
  {title} {\bibinfo {title} {Measurement-based quantum computation},\ }\href
  {https://doi.org/10.1038/nphys1157} {\bibfield  {journal} {\bibinfo
  {journal} {Nature Physics}\ }\textbf {\bibinfo {volume} {5}},\ \bibinfo
  {pages} {19} (\bibinfo {year} {2009})}\BibitemShut {NoStop}%
\bibitem [{\citenamefont {Raussendorf}\ and\ \citenamefont
  {Wei}(2012)}]{annurev020911125041}%
  \BibitemOpen
  \bibfield  {author} {\bibinfo {author} {\bibfnamefont {R.}~\bibnamefont
  {Raussendorf}}\ and\ \bibinfo {author} {\bibfnamefont {T.-C.}\ \bibnamefont
  {Wei}},\ }\bibfield  {title} {\bibinfo {title} {Quantum computation by local
  measurement},\ }\href
  {https://doi.org/https://doi.org/10.1146/annurev-conmatphys-020911-125041}
  {\bibfield  {journal} {\bibinfo  {journal} {Annual Review of Condensed Matter
  Physics}\ }\textbf {\bibinfo {volume} {3}},\ \bibinfo {pages} {239} (\bibinfo
  {year} {2012})}\BibitemShut {NoStop}%
\bibitem [{\citenamefont {Liu}\ \emph {et~al.}(2022)\citenamefont {Liu},
  \citenamefont {Zhou},\ and\ \citenamefont {Chen}}]{PhysRevB.106.144311}%
  \BibitemOpen
  \bibfield  {author} {\bibinfo {author} {\bibfnamefont {H.}~\bibnamefont
  {Liu}}, \bibinfo {author} {\bibfnamefont {T.}~\bibnamefont {Zhou}},\ and\
  \bibinfo {author} {\bibfnamefont {X.}~\bibnamefont {Chen}},\ }\bibfield
  {title} {\bibinfo {title} {Measurement-induced entanglement transition in a
  two-dimensional shallow circuit},\ }\href
  {https://doi.org/10.1103/PhysRevB.106.144311} {\bibfield  {journal} {\bibinfo
   {journal} {Phys. Rev. B}\ }\textbf {\bibinfo {volume} {106}},\ \bibinfo
  {pages} {144311} (\bibinfo {year} {2022})}\BibitemShut {NoStop}%
\bibitem [{\citenamefont {Guo}\ \emph {et~al.}(2023)\citenamefont {Guo},
  \citenamefont {Zhang}, \citenamefont {Bi},\ and\ \citenamefont
  {Yang}}]{PhysRevResearch.5.043069}%
  \BibitemOpen
  \bibfield  {author} {\bibinfo {author} {\bibfnamefont {Y.}~\bibnamefont
  {Guo}}, \bibinfo {author} {\bibfnamefont {J.-H.}\ \bibnamefont {Zhang}},
  \bibinfo {author} {\bibfnamefont {Z.}~\bibnamefont {Bi}},\ and\ \bibinfo
  {author} {\bibfnamefont {S.}~\bibnamefont {Yang}},\ }\bibfield  {title}
  {\bibinfo {title} {Triggering boundary phase transitions through bulk
  measurements in two-dimensional cluster states},\ }\href
  {https://doi.org/10.1103/PhysRevResearch.5.043069} {\bibfield  {journal}
  {\bibinfo  {journal} {Phys. Rev. Res.}\ }\textbf {\bibinfo {volume} {5}},\
  \bibinfo {pages} {043069} (\bibinfo {year} {2023})}\BibitemShut {NoStop}%
\bibitem [{\citenamefont {Liu}\ \emph {et~al.}(2025{\natexlab{c}})\citenamefont
  {Liu}, \citenamefont {Ravindranath},\ and\ \citenamefont
  {Chen}}]{PhysRevB.111.024312}%
  \BibitemOpen
  \bibfield  {author} {\bibinfo {author} {\bibfnamefont {H.}~\bibnamefont
  {Liu}}, \bibinfo {author} {\bibfnamefont {V.}~\bibnamefont {Ravindranath}},\
  and\ \bibinfo {author} {\bibfnamefont {X.}~\bibnamefont {Chen}},\ }\bibfield
  {title} {\bibinfo {title} {Quantum entanglement phase transitions and
  computational complexity: Insights from ising models},\ }\href
  {https://doi.org/10.1103/PhysRevB.111.024312} {\bibfield  {journal} {\bibinfo
   {journal} {Phys. Rev. B}\ }\textbf {\bibinfo {volume} {111}},\ \bibinfo
  {pages} {024312} (\bibinfo {year} {2025}{\natexlab{c}})}\BibitemShut
  {NoStop}%
\bibitem [{\citenamefont {Sukeno}\ and\ \citenamefont
  {Okuda}(2023)}]{SciPostPhys.14.5.129}%
  \BibitemOpen
  \bibfield  {author} {\bibinfo {author} {\bibfnamefont {H.}~\bibnamefont
  {Sukeno}}\ and\ \bibinfo {author} {\bibfnamefont {T.}~\bibnamefont {Okuda}},\
  }\bibfield  {title} {\bibinfo {title} {{Measurement-based quantum simulation
  of Abelian lattice gauge theories}},\ }\href
  {https://doi.org/10.21468/SciPostPhys.14.5.129} {\bibfield  {journal}
  {\bibinfo  {journal} {SciPost Phys.}\ }\textbf {\bibinfo {volume} {14}},\
  \bibinfo {pages} {129} (\bibinfo {year} {2023})}\BibitemShut {NoStop}%
\bibitem [{\citenamefont {Okuda}\ \emph {et~al.}(2024)\citenamefont {Okuda},
  \citenamefont {Parayil~Mana},\ and\ \citenamefont
  {Sukeno}}]{PhysRevResearch.6.043018}%
  \BibitemOpen
  \bibfield  {author} {\bibinfo {author} {\bibfnamefont {T.}~\bibnamefont
  {Okuda}}, \bibinfo {author} {\bibfnamefont {A.}~\bibnamefont
  {Parayil~Mana}},\ and\ \bibinfo {author} {\bibfnamefont {H.}~\bibnamefont
  {Sukeno}},\ }\bibfield  {title} {\bibinfo {title} {Anomaly inflow, dualities,
  and quantum simulation of abelian lattice gauge theories induced by
  measurements},\ }\href {https://doi.org/10.1103/PhysRevResearch.6.043018}
  {\bibfield  {journal} {\bibinfo  {journal} {Phys. Rev. Res.}\ }\textbf
  {\bibinfo {volume} {6}},\ \bibinfo {pages} {043018} (\bibinfo {year}
  {2024})}\BibitemShut {NoStop}%
\bibitem [{\citenamefont {Kuno}\ \emph {et~al.}(2024)\citenamefont {Kuno},
  \citenamefont {Orito},\ and\ \citenamefont {Ichinose}}]{PhysRevB.109.054432}%
  \BibitemOpen
  \bibfield  {author} {\bibinfo {author} {\bibfnamefont {Y.}~\bibnamefont
  {Kuno}}, \bibinfo {author} {\bibfnamefont {T.}~\bibnamefont {Orito}},\ and\
  \bibinfo {author} {\bibfnamefont {I.}~\bibnamefont {Ichinose}},\ }\bibfield
  {title} {\bibinfo {title} {Bulk-measurement-induced boundary phase transition
  in toric code and gauge higgs model},\ }\href
  {https://doi.org/10.1103/PhysRevB.109.054432} {\bibfield  {journal} {\bibinfo
   {journal} {Phys. Rev. B}\ }\textbf {\bibinfo {volume} {109}},\ \bibinfo
  {pages} {054432} (\bibinfo {year} {2024})}\BibitemShut {NoStop}%
\bibitem [{\citenamefont {Stollenwerk}\ and\ \citenamefont
  {Hadfield}(2024)}]{stollenwerk2024}%
  \BibitemOpen
  \bibfield  {author} {\bibinfo {author} {\bibfnamefont {T.}~\bibnamefont
  {Stollenwerk}}\ and\ \bibinfo {author} {\bibfnamefont {S.}~\bibnamefont
  {Hadfield}},\ }\href {https://arxiv.org/abs/2403.11514} {\bibinfo {title}
  {Measurement-based quantum approximate optimization}} (\bibinfo {year}
  {2024}),\ \Eprint {https://arxiv.org/abs/2403.11514} {arXiv:2403.11514
  [quant-ph]} \BibitemShut {NoStop}%
\bibitem [{\citenamefont {Chen}\ and\ \citenamefont {Zhu}(2026)}]{Zenodo_data}%
  \BibitemOpen
  \bibfield  {author} {\bibinfo {author} {\bibfnamefont {G.-H.}\ \bibnamefont
  {Chen}}\ and\ \bibinfo {author} {\bibfnamefont {G.-Y.}\ \bibnamefont {Zhu}},\
  }\bibfield  {title} {\bibinfo {title} {Data for "born-rule statistical
  dynamical quantum phase transitions under measurement"},\ }\href
  {https://doi.org/10.5281/zenodo.20174434} {10.5281/zenodo.20174434} (\bibinfo
  {year} {2026})\BibitemShut {NoStop}%
\end{thebibliography}%

\clearpage

\section{End Matter}\label{ End matter}

Analytical quench: The 1D transverse-field Ising model can be mapped to noninteracting fermions via the Jordan-Wigner transformation and diagonalized as $\sum_k \epsilon_h (k) \alpha_k^\dagger \alpha_k$ with $\epsilon_h(k) = 2J\sqrt{(h-\cos k)^2+\sin^2 k }$. The DQPT is detected by nonanalyticities of the post-selected $f(+\cdots+)$ in the thermodynamic limit. Its analytic expression under PBC is given by~\cite{PhysRevLett.101.120603,PhysRevB.87.195104} 
\begin{equation} \label{eqpbc}
	f(+\cdots+)= -2{\rm Re}\int^\pi_0 \frac{dk}{2\pi}\ln[\cos^2\phi_k +\sin^2\phi_k e^{-2i\epsilon_h (k)t}],
\end{equation}
with $\phi_k = -\theta_h(k)$, $\theta_h(k) = \arctan [\sin k/(h-\cos k)]/2$. The period of kinks is determined by $h$ as $t^*=\pi/2\sqrt{(1-h^2)}$. The zeros of Eq.~\ref{eqpbc} are located along the lines
\begin{equation} \label{eqzeros}
	z_m(k) = \frac{1}{2\epsilon_h(k)}[\ln (\tan^2 \phi_k) + i\pi(2m+1)],m\in\mathbb{Z}\ .
\end{equation}

\begin{figure}[htbp]
	\centering
	\includegraphics[width=\columnwidth]{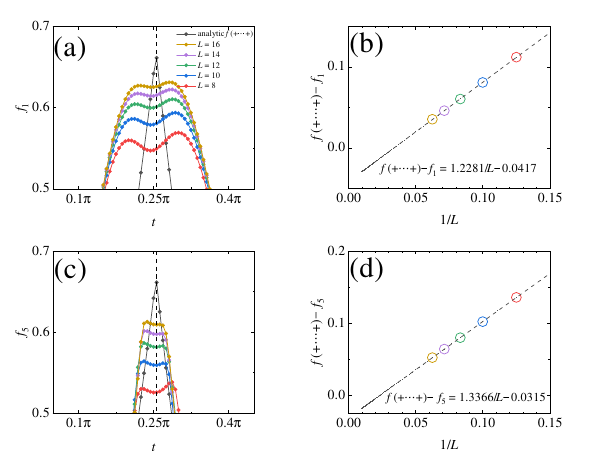}
	\caption{
	{\bf Finite-size level inversion.}
	Average dynamical free energy (a) $f_1$ and (c) $f_5$ in different system sizes around the first critical time with $h=0.2$ (indicated by dashed lines). The analytic $f(+\cdots+)$ (black) is obtained by Eq.~\ref{eqpbc}. Compared with $f_1$, high-moment $f_5$ showing a narrower half-height width lies closer to $f(+\cdots+)$. (b), (d): The difference between $f_n$ and analytic $f(+\cdots+)$ as a function of $1/L$ is labeled by corresponding colors with the fitted linear equation. This can be regard as the gap between the low temperature thermal state and paramagnetic state. Negative intercepts manifest the post-selected gap can be closed by finite but larger size as $L=1.2281/0.0417\approx30$ and $L=1.3366/0.0315\approx43$.
	}
	\label{fig:levelinversion} 
\end{figure}

Scaling of the average free energy: In Fig.~\ref{fig:levelinversion}, $f_n$ around the first critical time are illustrated with the analytic post-selected $f(+\cdots+)$ as a benchmark. The half-height width of these peaks gradually narrow as $L$ increases, approaching the sharp nonanalytic kink of $f(+\cdots+)$ in the thermodynamic limit. For higher moment $n=5$, the convergence toward the kink is significantly accelerated. The relation between system size $L$ and $f(+\cdots+)$ without field has been analytically given as $f(+\cdots+) = -2\ln[\cos^L (t)+\sin^L (t)]/L$, with local maxima at $t=\pi/4+m\pi/2\rightarrow |\cos(t)| = |\sin(t)| = 1/\sqrt{2}$. At these points, the relation between $f(+\cdots+)$ and $1/L$ is derived as 
\begin{equation} \label{eqlinear}
	f(+\cdots+) = \ln2-\frac{1}{L}\ln4\ .
\end{equation}
We therefore assume an analogous linear relation between $f_n$ at the critical time and $1/L$. The insets confirm the linear scaling of the deviation $f(+\cdots+)-f_n$ versus $1/L$ and the larger slope in Fig.~\ref{fig:levelinversion}(b) indicates the convergence. This demonstrates that increasing the moment $n$ effectively lowers the ``temperature" of the statistical ensemble, as a spectral filter that suppresses contributions from high-energy bitstrings and thus recovers the post-selected behavior. The fitted intercept approaching zero suggests that while the DQPT kinks are smoothed out by Born averaging at low moments or level crossing from some other low-energy states, the critical features can be recovered by increasing the system size $L$ with higher moments $n$. 

Additionally, there should be a critical $n$ where the DQPT can be observed by $f_n$. This critical $n$ is heuristically determined by the competition between the ground-state dominance and the thermal broadening of the ensemble. Qualitatively, if one wants to identify the nonanalytic kink through averaging $f_{n<\infty}$, the ground state must hold an absolute advantage compared with other levels, so that the weight factor of the ground state is much larger than others.
For a finite free energy gap $\Delta f=f_{\rm ex}-f_{\rm gs}$ between the first excited level and ground state at the critical time, the relative weight in the low-order approximation should satisfy the following relation: 
\begin{equation} \label{eqwei}
	\frac{P^n_{\rm ex}}{P^n_{\rm gs}}=\frac{e^{-nLf_{\rm ex}}}{e^{-nLf_{\rm gs}}}=e^{-nL\Delta f}\ll1 \ .
\end{equation}
Thus, the moment $n$ must greatly exceed $(L\Delta f)^{-1}$ to ensure the ground-state contribution dominates. However, from the analysis of the extended phase and level distribution, $\Delta f$ around the critical time is so small that larger $n$ and $L$ are required. In non-integrable systems with a continuous spectrum and extensive entanglement, the density of low-energy states may scale differently, potentially requiring other specific relations between $L$ and $n$.

\clearpage

\appendix

\section{Energy level distribution.} \label{Energy level distribution}

\begin{figure}[htbp]
\centering
\includegraphics[width=0.7\columnwidth]{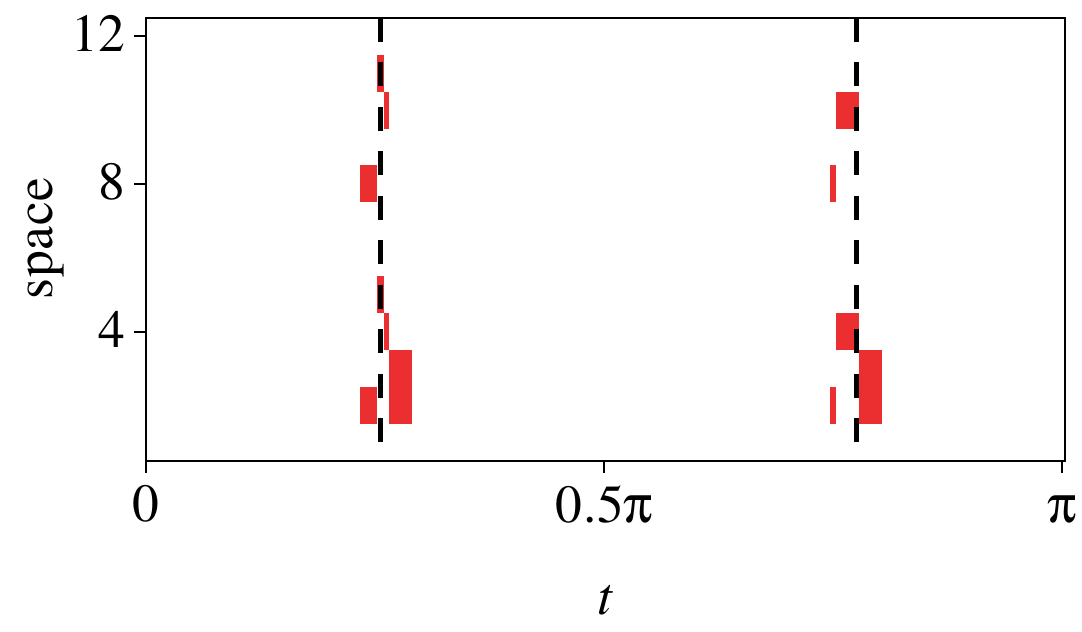}
\caption{
{\bf Ground-state trajectory.}
The bitstring of ground state is represented in the Pauli-$X$ basis at time $0\sim \pi$ and each pixel denotes $\ket{+}$ (white) or $\ket{-}$ (red) for one qubit at a particular time. The critical times are labeled by dashed lines.}
 \label{trajectory}
\end{figure}

We plot the ground-state trajectory in Fig.~\ref{trajectory} and the ground state is mostly occupied by $\ket{+\cdots+}$ except the points around critical times where some other bitstrings hold lower free energies, leading to the level crossing.

\begin{figure}[htbp]
\centering
\includegraphics[width=\columnwidth]{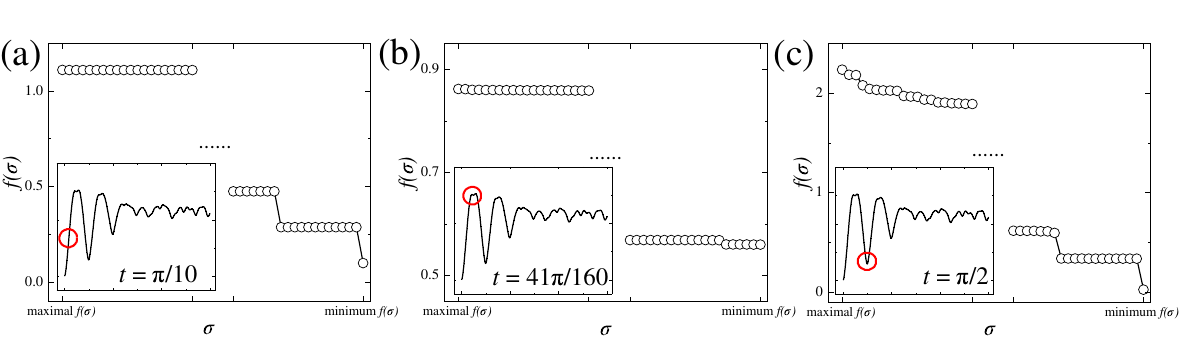}
\caption{
{\bf Spectrum at time slices.}
 Typical distribution of $f(\sigma)$ at three time slices ($t= \pi/10, 41\pi/160, \pi/2$). 20 highest and lowest energy levels are selected from the dynamics in Fig.~\ref{fig:fnband}(b). At early times (a) and (b), many bitstrings share nearly degenerate energy levels, but this degeneracy is visibly removed in (c). Inset: the $f_1$ with time slices labeled by red circles.}
 \label{f_distribution}
\end{figure}

The spectrum of single-shot free energies $f(\sigma)$ across representative bitstrings at given time slices are provided to show the extended phase. From the plots we roughly conclude that near the critical times, the energy levels of most bitstrings condense into a relatively narrower band ($0.5\sim 0.9$ in Fig.~\ref{f_distribution}(b)), so that the gap between the ground state and other excited states tends to close, leading to the extended phase and level crossings in finite-size systems. For slices far from critical times (Fig.~\ref{f_distribution}(a) and Fig.~\ref{f_distribution}(c)), the level distribution ranges are broadened to $0\sim1$ ($t=\pi/10$) and $0\sim2$ ($t=\pi/2$). The gap between $f(+\cdots+)$ and the second-lowest $f(\sigma)$ can be observed.

\section{Born sampling.} \label{Born sampling}

\begin{figure}[htbp]
\centering
\includegraphics[width=\columnwidth]{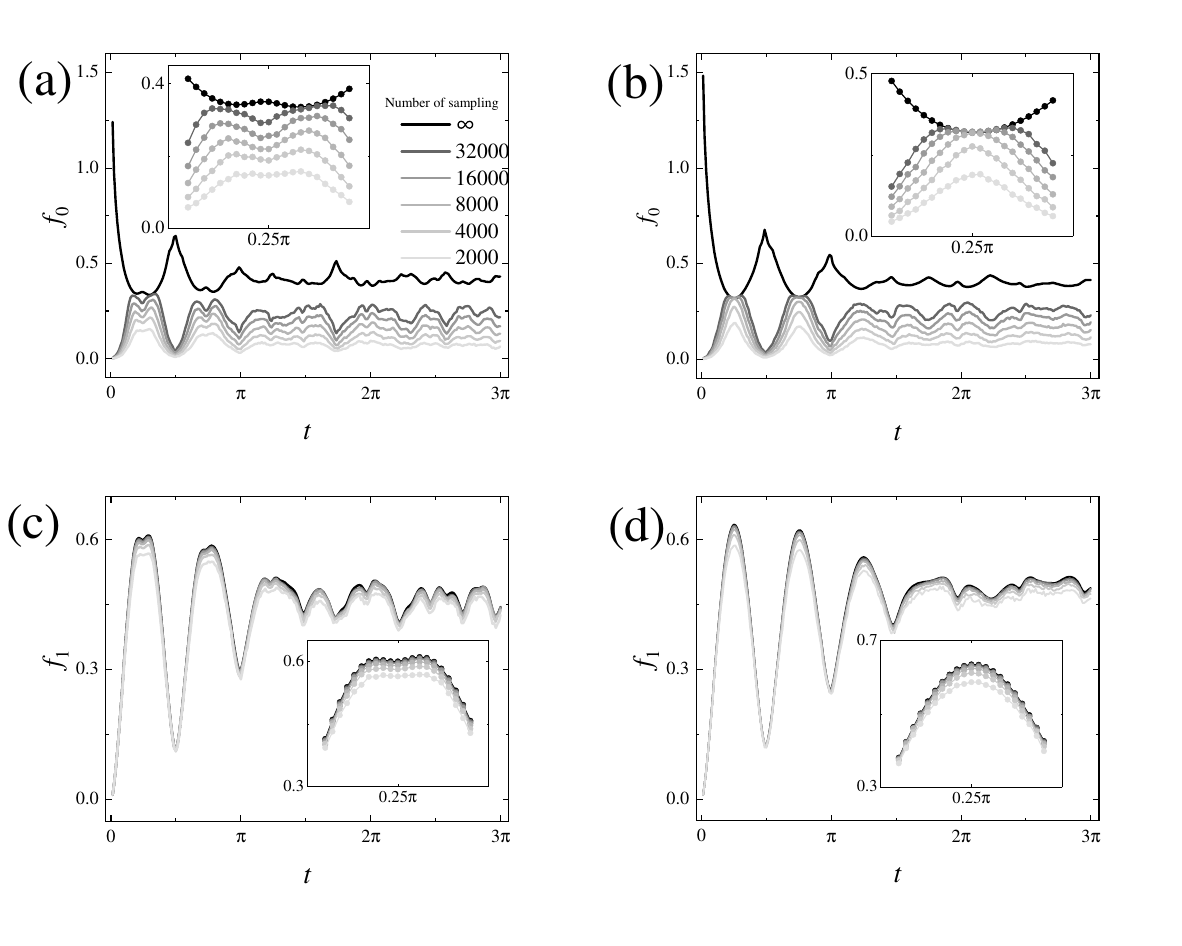}
\caption{
{\bf Numerical Born sampling} with $h = 0.2$.
Obtaining the probability $P(\sigma)$ in Eq.~\eqref{eq:mixed} and further $f_n$ through multiple samplings. Note that the resulting average of $\sigma$ may not include all possible bitstrings, because some high-energy $\sigma$ are unlikely to occur during sampling. The shading of curves is determined by the number of samples, where $\infty$ denotes the exact computation over all possible bitstrings. $f_0$: infinite-temperature average under (a) PBC and (b) OBC. $f_1$: unit-temperature average under (c) PBC and (d) OBC. Inset: more detailed data around $t=0.25\pi$.}
\label{fsample} 
\end{figure}

Here, we present the average dynamical free energy $f_n$ computed from a numerically simulated sampling process. For repeated Born measurements, the square of partition function $P(\sigma)$ can be regarded as the probability of a specific bitstring $\sigma$. 

We have computed $f_{0}$ and $f_1$ under two boundary conditions with the number of sampling ranging from 2000 to 32000, see Fig.\ref{fsample}. For $f_1$, sampling results are approaching the exact results with increasing the number of sampling. The characteristics of the exact computation are sufficient to be captured even with a relatively small number of sampling. 

However, some extremely large $f(\sigma)$ can dominate the equal-weighted summation and significantly increase $f_0$, but these bitstrings are almost impossible to occur for sampling outcomes. As a result, the sampling $f_0$ are completely different from the exact case, with their local extreme values being opposite. Based on the free energy spectrum and extended phase given in the main text, the $f(\sigma)$ of various bitstrings lie closer to each other at the critical times. Under this condition, increasing the number of sampling enables the emergence of a broader variety of bitstrings, agreeing better with the exact result. Therefore, $f_0$ around critical times exhibits an obvious enhancement and gradually converges to the exact value.

\section{Time slices in complex plane} \label{Time slices in complex plane}

Here, we provide the numerical details in the complex time plane at typical slices as supplementary data. 

\begin{figure}[htbp]
\centering
\includegraphics[width=\columnwidth]{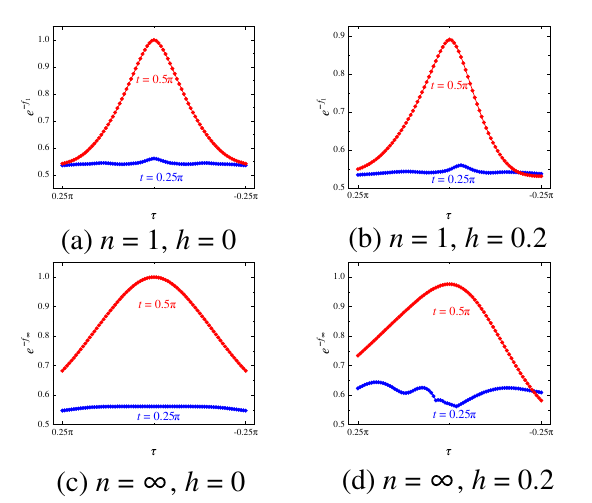}
\caption{
{\bf Real time slices.}
The imaginary-time evolution of $e^ {-f_{1 / \infty}}$ at fixed real time $t = 0.25\pi,\ 0.5\pi$ extracted from the complex time plane. The critical time for $h=0.2$ is slightly later than $t=0.25\pi$.}
\label{fixedt}
\end{figure}

\begin{figure}[htbp]
\centering
\includegraphics[width=\columnwidth]{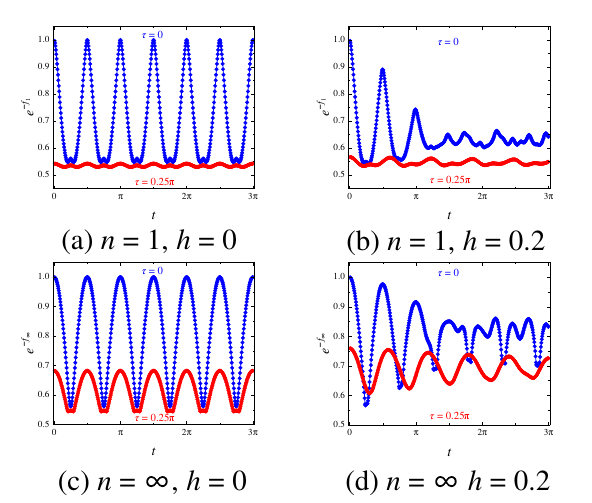}
\caption{
{\bf Imaginary time slices.}
The real-time evolution of $e^ {-f_{1 / \infty}}$ at fixed imaginary time $\tau = 0,\ 0.25\pi$ extracted from the complex time plane.}
\label{fixedtau} 
\end{figure}

At fixed real time shown in Fig.~\ref{fixedt}, the zeros of post-selected $e^{-f(\sigma)}$ around the critical time vanish under Born averaging and $e^{-f_{1}}$ exhibits slight fluctuations during the imaginary-time evolution. For $n = \infty$ in Fig.~\ref{fixedt}(c), the bitstring with lower energy around $e^{-f(\sigma)} = 0.55$ acts as the ground state and replaces high-energy zeros in Fig.~\ref{fig:complexpbc}(a), showing the uniform line $t=0.25\pi$ which is almost unchanged along the imaginary-time direction. Furthermore, the smoothed kinks due to the finite-size level crossing discussed in the main text can be observed in Fig.~\ref{fixedtau}.

\section{Dynamical free energy under OBC.} \label{Dynamical free energy under OBC}

\begin{figure*}[ht]
\centering
\includegraphics[width=0.9\textwidth]{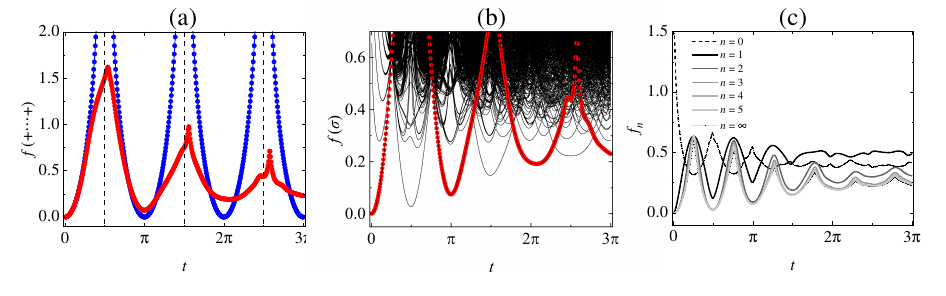}
\caption{
{\bf Free energy spectrum dynamics under OBC.} (a) The Dynamical free energy $f(+\cdots+)$ in finite size $L = 12$ under OBC. The nonanalytic period $t^*$ without transverse field is exactly $\pi$ labeled by vertical dashed lines. (b) The time dependent ground state undergoes spontaneous oscillations between the paramagnetic state $\ket{+\cdots+}$ and edge state $\ket{-+\cdots+-}$,  which ``energetically" favours $-$ i.e. a spin-flip on the two edges, showing the level crossing different from PBC. (c) Average free energy $f_n$ with moments $n=0\sim5$ and $\infty$.}
\label{fig:fnbando} 
\end{figure*}

\begin{figure}[ht]
\centering
\includegraphics[width=\columnwidth]{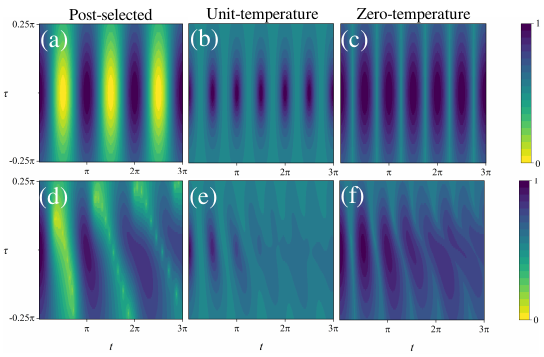}
\caption{
{\bf Complex time plane under OBC}. 
Finite-size evolution of $e^{-f(+\cdots+)} = P^{1/L}(+\cdots+)$ and $e^{-f_n}$ in the complex time plane under OBC with field $h = 0$ (upper panel) and $h=0.2$ (lower panel). (a), (d) Post-selected $e^{-f(+\cdots+)} $: Periodic zeros lie on the real axis but are absent in the complex plane $\tau \ne0$. (b), (e) $e^{f_1}$: The zeros on the real axis disappear and the real-time period is halved, similar to the results under PBC. (c), (f) $e^{f_\infty}$: The kinks from the crossing between the paramagnetic state and edge state, are extended throughout the entire plane. Compared with the smoother ground state of PBC, these sharp kinks only occur at a time point. 
}
\label{complexobc} 
\end{figure}

In this section, we supplement the numerical results of the dynamical free energy under OBC.

As shown in Fig.~\ref{fig:fnbando}(a), the period of kinks is doubled to $\pi$ and $f(+\cdots+)$ without field at critical times will diverge to infinity, corresponding to the exact zeros of partition function. Following the energy spectrum, the level crossing under OBC exhibits distinct behavior compared to PBC. This OBC ground state comprises two separate contributions from $f(+\cdots+)$ and $f(-+\cdots+-)$, where the lower one dominates the ground-state configuration and the switching between these two states gives rise to nonanalytic kinks with period $\pi/2$. Similar results are also observed for finite moment $f_{n}$ in Fig.~\ref{fig:fnbando}(c), where original kinks become continuous under low-moment averaging. It can be concluded that the average free energy of a finite system exhibits level crossings under both boundary conditions. For PBC, the level crossings arise from the finite-size effect and vanish as the system size increases, while the level crossing under OBC is from edge states independent of the size.

Extending above results to the partition function density in the complex time plane. In Fig.~\ref{complexobc}(a) with $h=0$, we can only observe isolated zeros with a period of $\pi$ on the real axis. These zeros disappear under the unit-temperature average due to their extremely low probability. For zero-temperature $n=\infty$, kinks from level crossing have become the lowest-energy configuration at $\tau\neq 0$, making them continuous and prominent lines across the entire plane.

\section{Measurement-based random circuits.} \label{Measurement-based random circuits}

When the Born-measurement outcome of every qubit happens to be positive ``0" (red arrows in the main text), this measurement protocol can be exactly mapped to the translationally invariant Hamiltonian evolution of Ising model. However, other outcomes with arbitrary ``1" (blue arrows) are likely to generate randomness into the unitary evolution, which inserts Pauli-$Z$ and $X$ between evolution operators in terms of the Pauli correction.
For the measurement-based circuits, there are three types of measurement basis in Fig.\ref{fmeasure} ($M(\theta,\phi)$ denotes the outcome of measurement basis on the Bloch sphere as $\cos{(\theta/2)}\ket{0}+ e^{i\phi}\sin(\theta/2)\ket{1}$): when the left edge qubit $(-2J, \pi/2)$ is measured to be ``1", two Pauli-$Z$ are inserted, 
\begin{equation} \label{eq2}
	M(-2J, \pi/2)=``1"\Rightarrow U = e^{ihX}Z_1Z_2e^{iJZZ}\ .
\end{equation}
When the vertex qubit $(\pi/2,0)$ is measured to be ``1", a Pauli-$Z$ is inserted,
\begin{equation} \label{eq3}
	M(\pi/2,0)_1=``1"\Rightarrow U = e^{ihX}Z_1e^{iJZZ}\ ,
\end{equation}
which is likely to break the Ising symmetry and generate the state with an odd number of $\ket{-}$. When the upper edge qubit $(\pi/2,2h)$ is measured to be ``1", a Pauli-$X$ is inserted,
\begin{equation} \label{eq4}
	M(\pi/2,2h)_1=``1"\Rightarrow U = e^{ihX}X_1e^{iJZZ}\ .
\end{equation}
An irrelevant global phase is omitted in the above derivation.

Moreover, due to the local entanglement structure of the cluster state where every bulk qubit is maximally entangled with its neighbors, all possible bulk measurement outcomes are equiprobable in this measurement-based unitary evolution protocol. That is, for an arbitrary outcome from the basis in Fig.\ref{fmeasure} $M = \left\{ 111\cdots,011\cdots \right\}$ of $N$ qubits in the bulk, probability $P(M) = 1/2^{N}$. Setting the total evolution operator decided by measurement outcomes $M$ as $U_t$. It is also unitary with inserted random Pauli operators. The probability for bitstring $\sigma$ at the boundary with bulk outcome $M$ can be expressed as
\begin{equation} \label{eqbulk}
	P(\sigma M) = \frac{1}{2^N}\bra{+\cdots+}U^\dagger_t \Pi_{\sigma}U_t\ket{+\cdots+}\ , 
\end{equation}
where $\Pi_{\sigma}$ is the projector of bitstring $\sigma$.

Consequently, if no correction is applied to the system to cancel out Pauli errors, the measurement-based evolution transforms the time evolution under a fixed Hamiltonian into a random circuit containing errors, which are equivalent to imposing pulses with $\pi$-phase rotations on qubits. However, for each bulk qubit, a $1/2$ probability of obtaining a measurement error causes the pulse to have a much greater effect on the circuit, while the Ising evolution can instead be regarded as a perturbation. The effect of pulse is positively correlated with the circuit depth (or height of the cylindrical cluster state) determined by $\Delta t$. The resulting random circuit can be viewed as a high-temperature limit of the Born ensemble, where the effective temperature scales with the error rate per Trotter step.

In the early stage, the direct $Z$ error will rotate the evolved state to be almost orthogonal to the initial $\ket{+\cdots+}$, leading to extremely high free energy. As the Ising evolution operators $U(\Delta t)$ act continuously, the overlap between the evolved state and the initial state tends to gradually increase. When the $\pi$-phase error and the accumulated Ising evolution balance each other, $f(+\cdots+)$ will decrease to a stable value even in the absence of external correction. As a result, the DQPT depending on the Ising interactions cannot be observed in the absence of Pauli correction. Furthermore, the randomness introduced by the above measurements can also be viewed as decoherence at an effective temperature (different from the temperature of $f_n$), which is proportional to the error rate. Sufficient Pauli corrections can lower the effective temperature to recover the DQPT.

\end{document}